%% file: Microring.tex
\documentclass[reprint, amsmath,amssymb,aps,twocolumn]{revtex4}

\usepackage{graphicx}
\usepackage{dcolumn}
\usepackage{bm}
\usepackage{gensymb}

\newcommand{\siot}{SiO\textsubscript{2}}
\newcommand{\sitnf}{Si\textsubscript{3}N\textsubscript{4}}
\newcommand{\chft}{CHF\textsubscript{3}}
\newcommand{\ot}{O\textsubscript{2}}

\begin{document}


\title{Microring resonators on a suspended membrane circuit for atom-light interactions}
\author{Tzu-Han Chang$^{ 1 }$, Brian Fields$^1$, May E. Kim$^{1,\dag}$\footnotetext{\small $^{\dag}$ Current address: National Institute of Standards and Technology, 325 Broadway, Boulder, CO 80305.}, and Chen-Lung Hung$^{1, 2, 3 ,\ast}$\footnotetext{\small $^{\ast}$ e-mail: clhung@purdue.edu.}}
\address{$^1$ Department of Physics and Astronomy, Purdue University, West Lafayette, IN 47907}
\address{$^2$ Purdue Quantum Science and Engineering Institute, Purdue University, West Lafayette, IN 47907}
\address{$^3$ Birck Nanotechnology Center, Purdue University, West Lafayette, IN 47907}

\date{\today}

\begin{abstract}
Creating strong coupling between quantum emitters and a high-fidelity photonic platform has been a central mission in the fields of quantum optics and quantum photonics. Here, we describe the design and fabrication of a scalable atom-light photonic interface based on a silicon nitride microring resonator on a transparent silicon dioxide-nitride multi-layer membrane. This new photonic platform is fully compatible with freespace cold atom laser cooling, stable trapping, and sorting at around 100 nm from the microring surface, permitting the formation of an organized, strongly interacting atom-photonic hybrid lattice. We demonstrate small radius (around 16~$\bm{\mu m}$) microring and racetrack resonators with a high quality factor ($\bm{Q}$) of $\bm{3.2\times 10^5}$, projecting a single atom cooperativity parameter ($\bm{C}$) of 25 and a vacuum Rabi frequency ($\bm{2g}$) of $\bm{2\pi\times 340}~$MHz for trapped cesium atoms interacting with a microring resonator mode. We show that the quality factor is currently limited by the surface roughness of the multi-layer membrane, grown using low pressure chemical vapor deposition (LPCVD) processes. We discuss possible further improvements to a quality factor above $\bm{5\times10^6}$, potentially achieving single atom cooperativity parameter higher than 500 for strong single atom-photon coupling. Our microring platform may also find applications in on-chip solid-state quantum photonics.
\end{abstract}

\maketitle

\section{Introduction}
Creating efficient atom-light nanophotonic interfaces with stably trapped atoms in their optical near-field can lead to a wide range of applications in quantum optics, quantum communications, and quantum many-body physics \cite{o2009photonic,Cirac_photonics_2017, ChangRMP2018}. Optical nanofibers \cite{vetsch_optical_2010,goban_demonstration_2012,kato_strong_2015, sorensen_coherent_2016,corzo_large_2016}, photonic crystal waveguides \cite{goban_superradiance_2015}, and cavities \cite{thompson_coupling_2013,tiecke_nanophotonic_2014} are exemplary platforms that have recently demonstrated atom trapping and large atom-light interactions in the evanescent field of their guided modes. Key enabling features for enhanced coupling are sub-diffraction transverse confinement of guided photons and enhanced photonic density of states, achieved either through forming micro- or nano-scale Fabry-Perot cavities or through slow light effects in photonic crystal waveguides. They permit several far-off resonant optical trapping schemes \cite{grimm2000optical} in the near-field, such as two-color evanescent field traps \cite{le_kien_atom_2004,balykin2004atom,lacroute2012state}, side-illuminating optical traps \cite{thompson_coupling_2013,goban_superradiance_2015,Rios_molecule_2017}, or a hybrid trap formed by Casimir-Polder vacuum force and a single-color optical potential in photonic crystal waveguides \cite{hung_trapped_2013,gonzalez-tudela_subwavelength_2015}. With near field ($\sim$100~nm) trapping above the dielectric surface, it is generally expected that tens to more than a hundred-fold increase in atom-photon coupling rate may be found for a trapped atom in a nanophotonic platform compared to those realized in mirror-based optical cavities and resonators. 

\begin{figure*}[t]
\centering
\includegraphics[width=1.5\columnwidth]{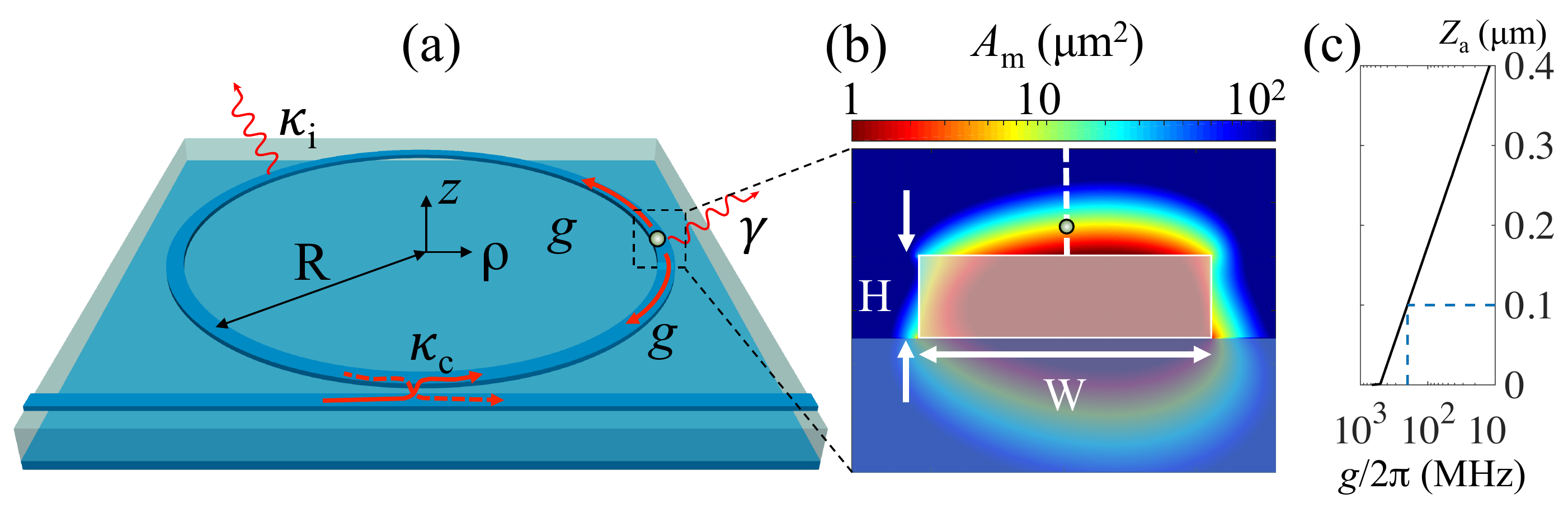}
\caption{Interfacing single atoms with a microring structure on a membrane for strong atom-light interactions. (a) Schematics of a silicon nitride microring (radius $R$) on a dioxide-nitride membrane, with a single trapped atom (green sphere). Curved arrows depict single atom-photon coupling  rate, $g$, to the resonator modes. Wavy arrows depict intrinsic resonator loss (at rate $\kappa_\mathrm{i}$) and the atomic decay (at rate $\gamma$), respectively. A linear bus waveguide couples to and from the resonator modes at rates $\kappa_c$ (depicted by crossed solid and dashed arrows). (b) The effective mode area $A_\mathrm{m}$ of a microring of width $W=1.1~\mu$m, height $H=0.29~\mu$m, and radius $R\sim$16$\mu$m; $A_\mathrm{m} \approx 5.2~\mu$m$^2$ at the depicted atom location $(\rho_\mathrm{a},z_\mathrm{a})=(R,100~$nm). Shaded structures mark the microring waveguide (\sitnf) and supporting membrane (\siot~and \sitnf ~layers), respectively. (c) Atom-photon coupling strength $g(z)$ along a vertical dashed line in (b); $g(z_\mathrm{a})/2\pi = 200~$MHz is marked by the dotted lines.}
\label{fig1}
\end{figure*}

On the other hand, achieving coherent quantum operations with high-fidelity requires ultra-low optical loss in the host dielectric nanostructures \cite{douglas_quantum_2015,gonzalez-tudela_subwavelength_2015}, and has remained a challenge. In cavity quantum electrodynamics (QED), the figure of merit for quantum coherence is favored by a large single atom cooperativity parameter $C= 4g^2/\kappa\gamma \gg 1$, requiring that the atom-photon coupling strength $g$ is large compared to the geometric mean of the photonic loss rate ($\kappa$) and the atomic radiative decay rate into freespace ($\gamma$). For a perfect quantum emitter, $g=\sqrt{\frac{3\lambda^3 \omega_0 \gamma}{16\pi^2V_\mathrm{m}}}$ (assuming a spherical symmetric dipole) and $C\sim \frac{3\lambda^3}{4\pi^2}\frac{  Q }{ V_\mathrm{m}}$ depends solely on the ratio between the quality factor $Q= \omega_0/\kappa$ of the photonic mode (of frequency $\omega_0$ and freespace wavelength $\lambda$) and the effective mode volume $V_\mathrm{m}$, which is inversely proportional to the guided-mode photon energy density at the atomic trap location. State-of-the-art high-Q nanophotonic platforms have yet to achieve cooperativity parameters $C \gg 10$ due to limited $Q/V_\mathrm{m}$ that stems mostly from fabrication imperfections. For example, $Q\sim 10^4$ and $V_\mathrm{m}/\lambda^3\sim 100$ is reported recently for a photonic crystal cavity \cite{tiecke_nanophotonic_2014} and an atom-induced cavity near the band edge of a photonic crystal waveguide \cite{hood2016atom,douglas_quantum_2015} that are tailored for coupling with alkali atoms at resonant wavelengths ranging from $\lambda \approx 780~$nm at rubidium D2-line to $894$~nm at cesium D1-line. There are multiple schemes in existence for boosting $Q/V_\mathrm{m}$ in atom-nanophotonic platforms \cite{ChangRMP2018}. For comparison, monolithic micro-photonic resonators such as silica micro-toroids \cite{aoki2006observation}, bottles \cite{o2013fiber}, and spheres \cite{shomroni2014all} or silicon nitride micro-disks \cite{barclay2006integration} are other photonic resonator structures with ultrahigh quality factors $Q>10^6$ but with large $V_\mathrm{m}/\lambda^3 >10^4$ for transit atoms in time-of-flight. The prospect of direct laser cooling and atom trapping in their optical near field has remained elusive, partially due to geometrical constraints and limited optical access through the dielectric structures and their substrates \cite{alton2013interacting}, if present.

In this paper, we report a planar-type microring photonic structure and its racetrack-variant design capable of simultaneous realizations of strong confinement for large atom-photon coupling rate, experimentally accessible atom trapping and sorting schemes, and high $Q/V_\mathrm{m}$, serving as a coherent, scalable atom-photon quantum interface. Our implementation is based on silicon nitride microring resonators that have recently achieved ultrahigh quality factors $Q >10^7$ in telecom optical wavelength band \cite{Xuan:16,Ji:17} and $Q > 3\times10^6$ close to Cs atomic spectroscopy bands \cite{Kaufmann:18}. We adapt the design to geometries tailored for cavity QED with neutral atoms and address major challenges that need to be overcome. We enable full optical access for laser cooling and trapping of alkali atoms, e.g. atomic cesium, directly on a microring by fabricating it on a transparent membrane substrate. Moreover, we investigate various trapping schemes including two-color evanescent field trap and top-illuminating optical tweezer trap at tunable distances around 100~nm above a resonator waveguide, as well as the combination of both schemes for atom array sorting.

\section{Overview of the platform}
\label{overview}
Figure 1 shows the schematics of our resonator platform. The microrings are fabricated on top of a suspended \siot-\sitnf~multi-layer membrane, formed by a $\sim 2$ $\mu$m-thick \siot~(silicon dioxide) layer and a $\sim 550$ nm-thick \sitnf~(silicon nitride) bottom-layer that can provide high tensile stress after being released from a silicon substrate to form a large window around an area of $\sim 2~$mm $\times$8~mm; see Fig.~2. The high tensile stress offered by the nitride bottom layer is necessary to preserve the optical flatness of the membrane. The transparent membrane allows laser beams to be sent from either top or bottom sides of the microring structure, allowing cold atoms to be directly laser cooled, trapped, and transported on the surface of a microring resonator \cite{kim2019}.

Due to its higher mode field intensity above the surface of the resonator waveguide (Supplement 1 Sec. 1-A), we utilize the fundamental transverse-magnetic (TM) mode for creating atom-light coupling. The cross section of the resonator waveguide is chosen for sufficient evanescent field strength above the waveguide surface while maintaining high $Q$. The small radius of the microring $R=16~\mu$m ensures a moderately small mode volume $V_\mathrm{m} = A_\mathrm{m}L$, where $L=2\pi R$ is the circumference of the ring and $A_\mathrm{m}$ is the effective mode area defined as
\begin{equation}
A_\mathrm{m}(\rho_\mathrm{a},z_\mathrm{a}) = \frac{\int \epsilon(\rho',z')|\mathbf{E}(\rho',z')|^2 d\rho'dz'}{\epsilon(\rho_\mathrm{a},z_\mathrm{a})|\mathbf{E}(\rho_\mathrm{a},z_\mathrm{a})|^2}.\label{eq:Am}
\end{equation}
Here $(\rho_\mathrm{a},z_\mathrm{a})$ denotes the transverse atomic location in cylindrical coordinates, $\epsilon$ is the dielectric function of the microring structure, and $\mathbf{E}$ is the TM-mode electric field. Figure~\ref{fig1} (b) plots the cross section of the effective mode area of a TM-mode at cesium D1-line $\lambda_\mathrm{D1}=894~$nm. A moderately small mode area $A_\mathrm{m}(R,z_\mathrm{a}) \approx 5.2~\mu$m$^2$ can be achieved when an atom is placed at around $z_\mathrm{a}\sim100~$nm above the microring surface, projecting a mode volume of $V_\mathrm{m} \approx 523~\mu$m$^3$, single-photon vacuum Rabi frequency $2g = 2\pi \times 400~$MHz, and a cooperativity parameter $C \sim 1\times 10^{-4}Q$. Achieving high $Q >10^6$ in such a small microring can thus make this platform well-suited for on-chip cavity QED experiments with high fidelity.

A linear bus waveguide is fabricated next to an array of microrings to couple to the clockwise (CW) and counter-clockwise (CCW) resonator modes. Away from the microring coupling region, the bus waveguide is tapered and extends all the way towards the edge of the transparent window where the waveguide is then embedded in a dioxide (or vacuum) top-cladding layer. 

As shown in Fig.~2, a U-shaped fiber groove is fabricated for epoxy fixture of a lensed optical fiber, which is edge-coupled to the bus waveguide with $\sim 70~$\% (or $50~$\% with vacuum cladding) single-pass coupling efficiency as expected through our finite-difference-time-domain (FDTD) calculations \cite{cite:lumerical}. We have currently achieved $\sim 30~$\% coupling efficiency with vacuum cladding. The lensed fiber, the edge-coupled bus waveguide, and an array of coupled microrings form a complete package of high-fidelity atom-light nanophotonics interface; see Fig. 2. 

\begin{figure}[t]
\centering
\includegraphics[width=0.9\columnwidth]{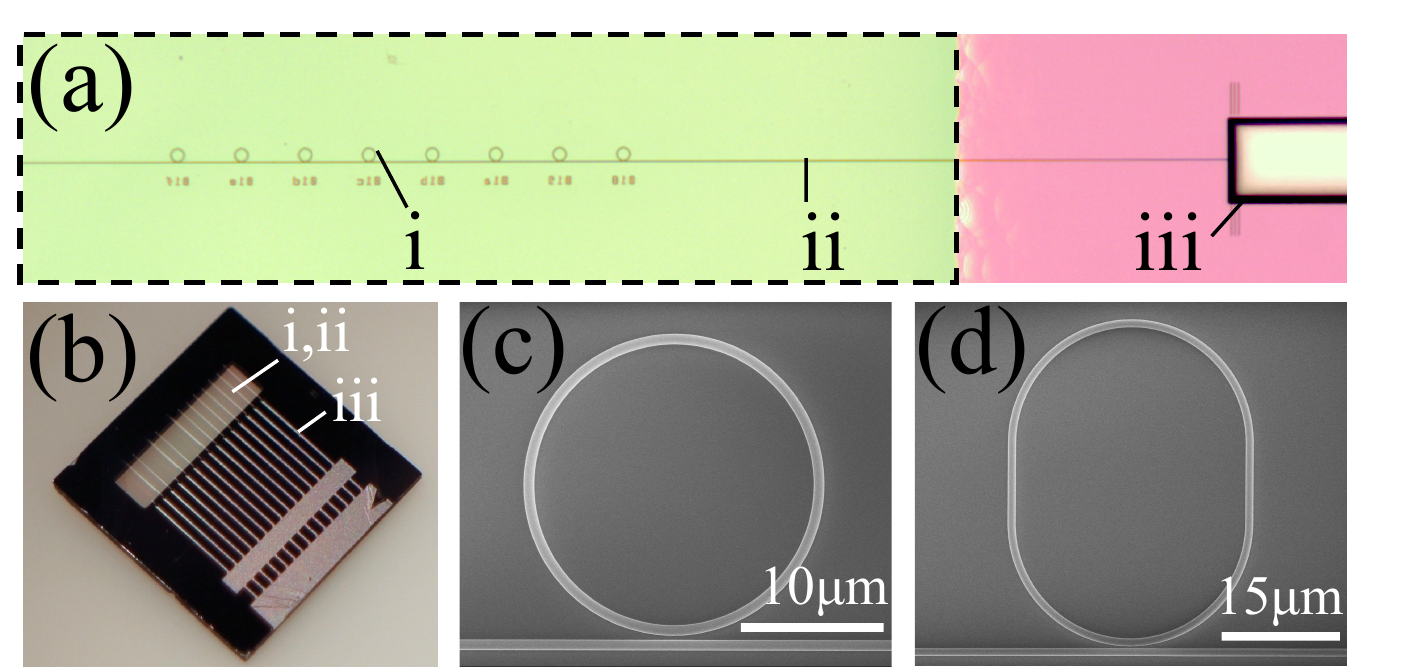}\\
\includegraphics[width=0.8\columnwidth]{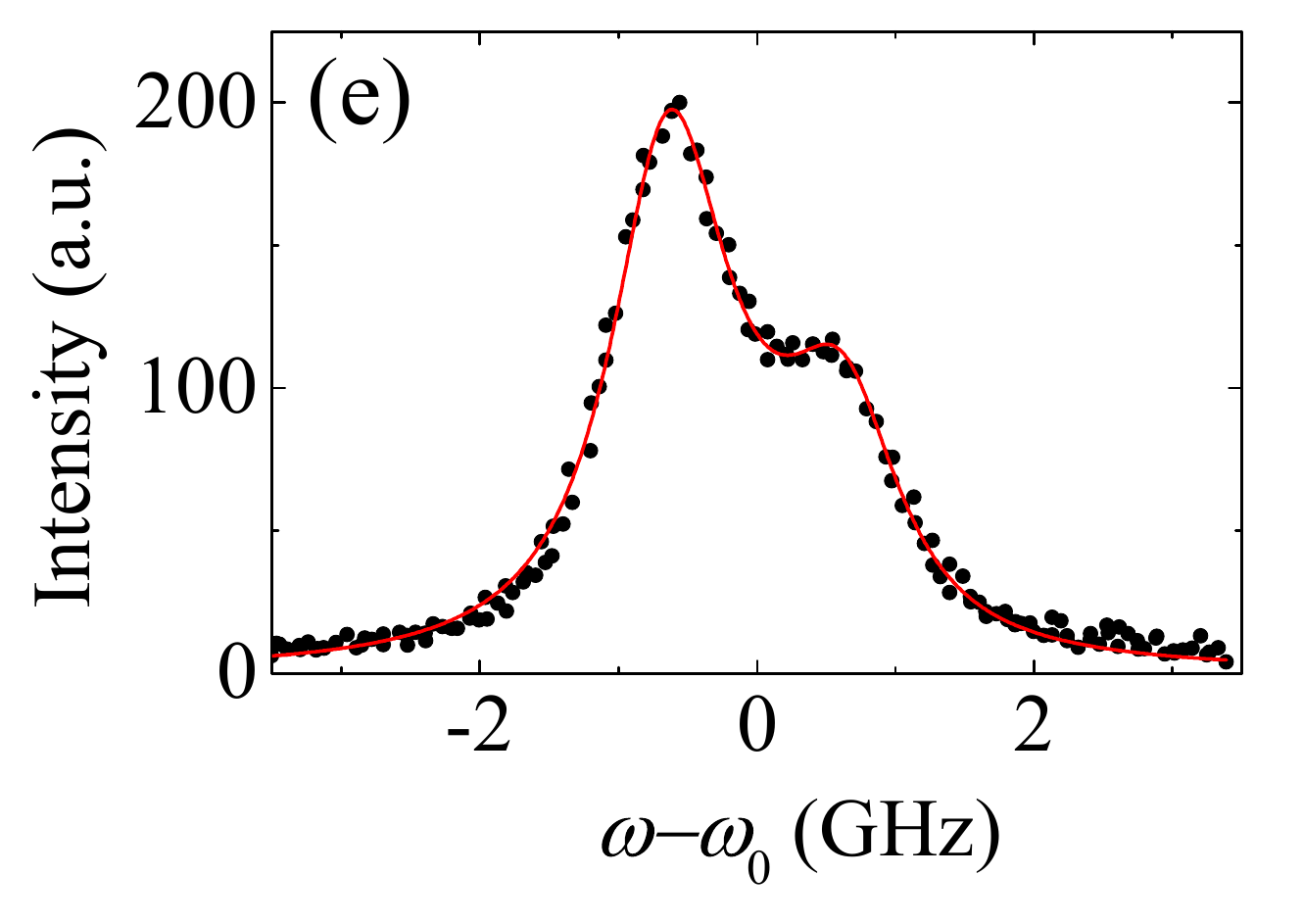}
\caption{Fabricated small radius microring/racetrack resonators and optical quality measurements. (a) Optical image of an array of microrings (i) coupled to a linear waveguide bus (ii) for fiber edge coupling in a U-groove (iii). Membrane area is enclosed in a dashed box. (b) Overview of the optical chip. The membrane is suspended within a 2~mm $\times$ 8~mm window. (c, d) Scanning electron micrographs (SEM) of fabricated (c) microring and (d) racetrack resonators, both with width $W=0.95~\mu$m and height $H=0.36~\mu$m. (e) Scattering intensity measurements near the resonance of a racetrack resonator. Solid line is a fit, giving $(\kappa,\beta,\omega_0)/2\pi= (1.01,0.655,334.792\times10^3)~$GHz.}
\label{fig:fab}
\end{figure}
\begin{figure}[b]
\centering
\includegraphics[width=1\columnwidth]{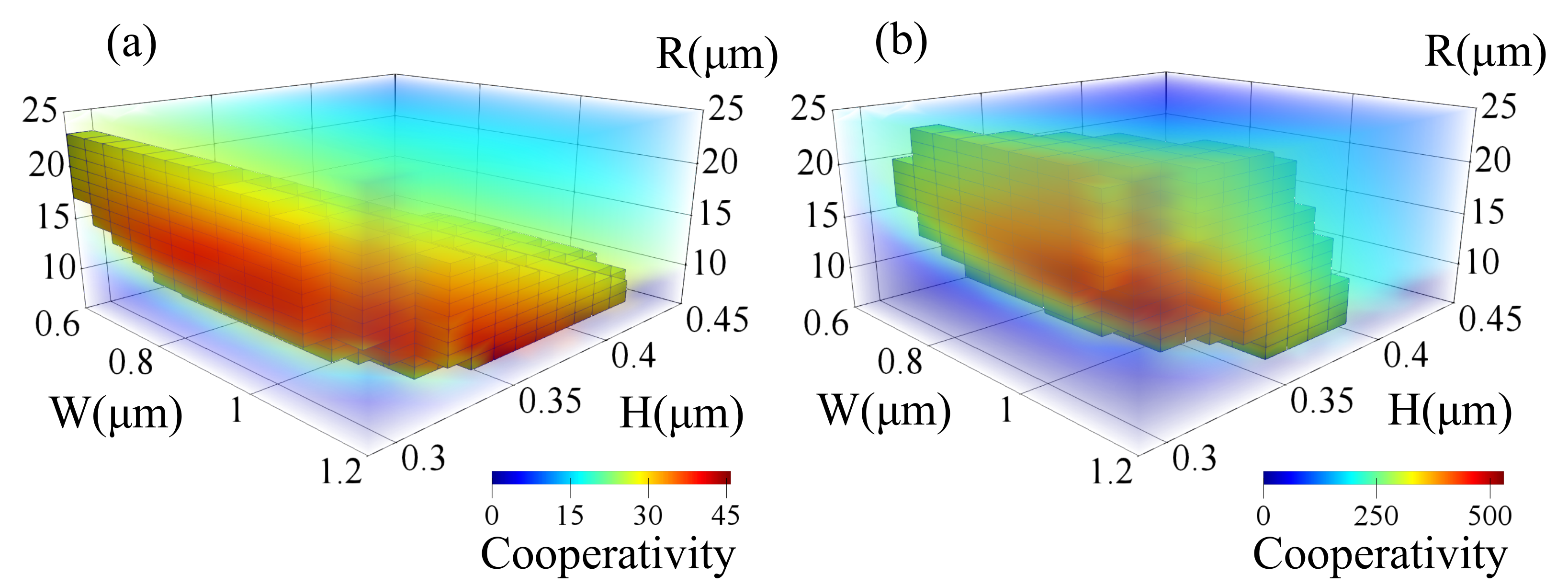}
\caption{Cooperativity optimization via scanning the microring geometry, with the surface roughness parameters ($\sigma_\pm$,$L_\pm$,$\sigma_\mathrm{t}$,$L_\mathrm{t}$,$\sigma_\mathrm{b}$,$L_\mathrm{b}$)= (a) (2,60,1.4,73,1.6,84)~nm and (b) (1.4,39,0.1,10,0.1,10)~nm, respectively. }
\label{fig:QoverV}
\end{figure}

\section{Fabrication of microring membrane  circuit and optical measurements}
\label{fab}
Figure 2 (a-b) show the optical image of a fabricated membrane optical circuit (see \cite{kim2019} for fabrication procedures). SEM images of microring and racetrack resonators on the membrane with coupling waveguide buses are shown in Fig. 2(c-d).

We characterize the quality factors near cesium D1 line by scanning the frequency $\omega$ of the coupled TM-mode and image the scattered light from individual rings on a charge-coupled-device (CCD) camera. The resonant frequency of the microring, $\omega_0$, has been thermally tuned by a freespace laser beam heating the silicon part of the optical circuit in vacuum \cite{tiecke_nanophotonic_2014}. Figure \ref{fig:fab}(d) shows a sample measurement. Double resonant peaks have been observed due to coherent back-scattering effect from fabrication imperfections that mixes the CW and CCW modes and creates an energy splitting (see Supplement 1 Sec.~1-B). Our measured CCD counts can be well fitted by a coupled-mode model \cite{Srinivasan2007} that captures mode-splitting and the peak asymmetry (see also Eq.~\ref{eq:elattice} and Supplement 1 Sec.~1-B). 
The fit gives total photon loss rate $\kappa/2\pi =$1.01~GHz, corresponding to an under-coupled quality factor of $Q\approx 3.2\times 10^5$ due to waveguide coupling rate $\kappa_c$ smaller than the intrinsic loss rate $\kappa_\mathrm{i}$. 

Using the measurement results and the fabricated geometry $(W,H)=(0.95,0.36)~\mu$m, we project the single atom cooperativity parameter to be $C \approx 25$, calculated using $(g, \kappa, \gamma)/2\pi \approx (170, 1010, 4.6 )~$MHz; under the same $Q$ and the geometry presented in Fig.~\ref{fig1}, we project $C\approx 35$ with $g/2\pi=200~$MHz. We note that there is still much room for improvement. Below we discuss in detail the optical loss analysis and $Q/V_\mathrm{m}$ optimization for maximizing cooperativity $C$.

\begin{figure*}[ht]
\centering
\includegraphics[width=1.8\columnwidth]{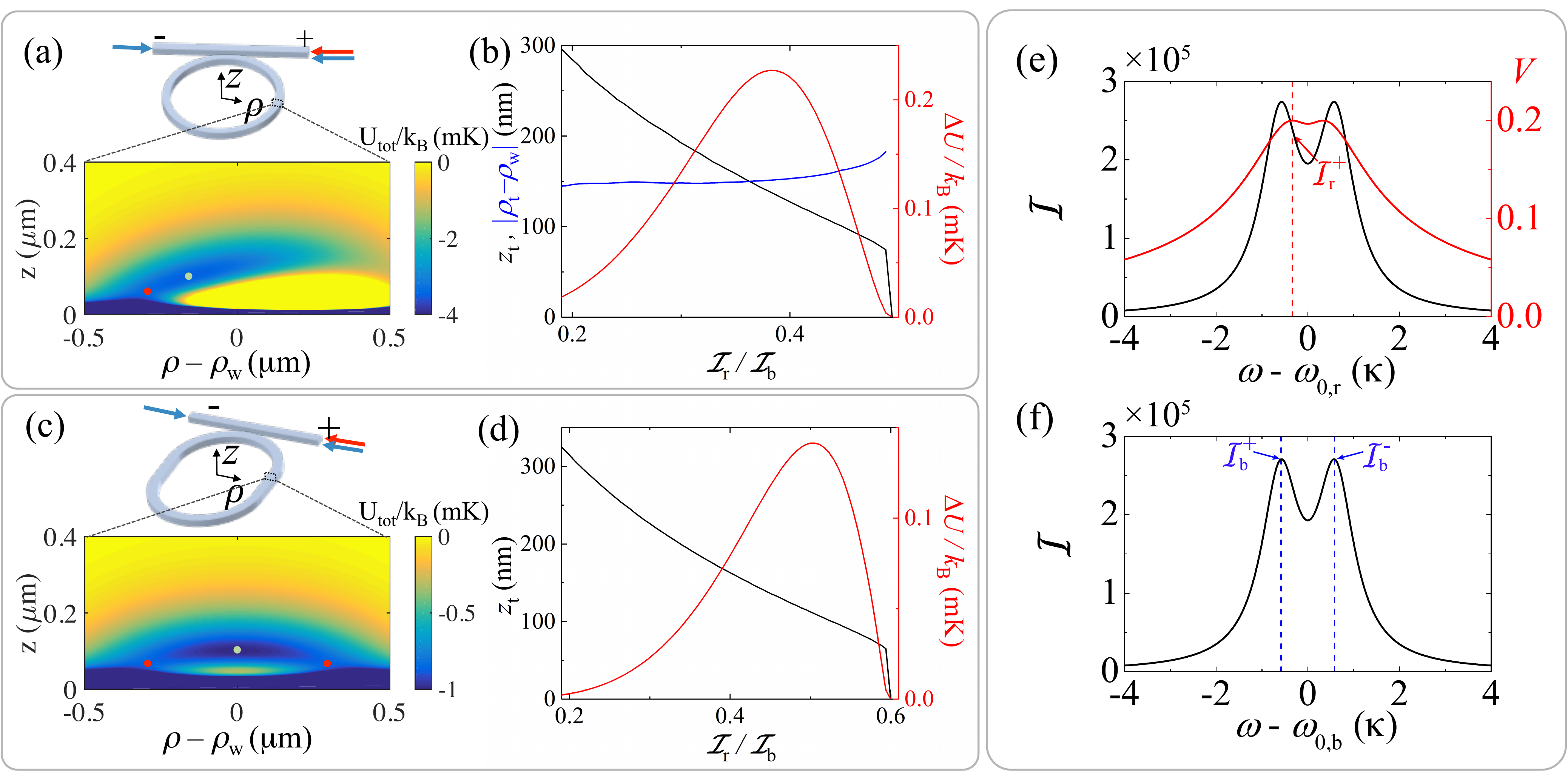}
\caption{Two-color evanescent field trap. (a,c) The coupling schemes are schematically shown in (a) for a microring and (c) for a racetrack resonator, respectively, where the injected lights are marked by red ($\omega_\mathrm{r}$) and blue ($\omega_\mathrm{b}$) arrows and the $\pm$ signs mark the direction of coupling. Sample total potential cross section $U_\mathrm{tot}(\rho,0,z)$ in the near-field region above the resonators (enclosed by dashed boxes) are displayed accordingly, where the top surfaces of the resonator waveguides are centered at $(\rho,z)=(\rho_\mathrm{w},0)$. Green spheres indicate the trap center and red spheres mark the positions of potential saddle points beyond which the trap opens. (b,d) Trap depth $\Delta U$ (red curves) and the vertical trap position $z_\mathrm{t}$ (black curves) can be adjusted by tuning the ratio of energy build up factors $\mathcal{I}_\mathrm{r}/\mathcal{I}_\mathrm{b}$ between the $\omega_\mathrm{r}$ and $\omega_\mathrm{b}$ modes; $\mathcal{I}_\mathrm{b}=$ (b) $5.4\times 10^5$, (d) $8.0\times 10^4$. For a microring trap (b), radial trap position $|\rho_\mathrm{t}-\rho_\mathrm{w}|$ (blue curve) remains roughly unchanged until the trap completely opens; for a racetrack trap (d), $\rho_\mathrm{t}=\rho_\mathrm{w}$. (e, f) Injected light frequencies and build up factors $\mathcal{I}$ (black curves) around the microring resonances. In (e), $\omega_\mathrm{r}$ (red dashed line) is chosen to maximize the visibility $V$ (red curve) and to eliminate the vector light shift (Supplement 1 Sec.~2). In (f), $\omega_\mathrm{b}$ modes (blue dashed line) are symmetrically excited around $\omega_\mathrm{0,b}$ to maximize $\mathcal{I}_\mathrm{b}^\pm $ and to eliminate potential corrugation and vector light shifts. Parameters used in (e,f): $(\kappa, \kappa_c, \beta)=2\pi \times (1, 0.5, 0.6)~$GHz.}
\label{fig:eft_xz}
\end{figure*}

\subsection{Current fabrication limit and mitigation methods}
Currently, surface scattering dominates the photon loss in our fabricated microrings; see also Supplement 1 Sec.~3-A for fundamental limits of the microring platform regarding material absorption. We have characterized the surfaces of the multi-layer film using atomic force microscopy (AFM) and obtained the root-mean-squared roughness $\sigma_\mathrm{t} \approx 1.4$ nm and the correlation length $L_\mathrm{t} \approx 73$~nm for the top nitride layer. For the bottom surface roughness of the microring, we infer from the surface quality of the dioxide middle layer, which we measured $(\sigma_\mathrm{b},L_\mathrm{b})\approx(1.6,84)~$nm. We estimate the edge roughness and correlation length to be around $(\sigma_\pm,L_\pm)\approx(2,60)~$nm by employing multipass e-beam writing technique \cite{Ji:17,Roberts:17} and optimized inductively coupled-plasma reactive-ion etching process with \chft/\ot~gas chemistry~\cite{Kruckel:15,Ji:17,Roberts:17}. In Supplement 1 Sec.~3-B, we adopt a volume current method to model the scattering loss rate due to the measured surface roughness \cite{Borselli:05}. Our result indicates that $Q_\mathrm{sim} \approx 3.5\times 10^5$ is in reasonable agreement with our measured quality factor.

Due to the major roughness incurred in the LPCVD-deposited dioxide layer, we note that the surface roughness of the microring is around three times worse than a typical single layer nitride deposited on a silicon wafer or on a thermally grown dioxide film. Possible improvements can be made by using a chemical mechanical polishing (CMP) technique to reduce the surface roughness in the top nitride layer and the middle dioxide layer as well. It has been reported that the surface roughness and the correlation length of a nitride~thin film can be greatly reduced down to $\sigma_\mathrm{t,b} \approx 0.1~$nm and $L_\mathrm{t,b} \approx 10~$nm from its original rough surface \cite{Ji:17}. Alternatively, the edge roughness and correlation length may be reduced to $\sigma_\pm \approx 1.39~$nm and $L_\pm \approx 39~$nm by using a plasma-assisted resist reflow technique \cite{Porkolab:14}. These immediate technological improvements permit a potential 10-fold increase in $Q/V_\mathrm{m}$, as will be discussed below.

\subsection{Q/V\textsubscript{m} optimization}
Given the characteristics of the surface quality and edge roughness, we perform finite element method (FEM) analysis \cite{cite:comsol} to obtain a geometrical design that maximizes $Q/V_\mathrm{m}$, concerning the dominant losses including the surface scattering loss and the waveguide bending loss. By scanning the cross-section and the radius of the microring, it is observed that the waveguide cross-section cannot be reduced indefinitely due to the constraint of surface scattering. Similarly, the radius $R$ of the ring is constrained to be above $\gtrsim 10~\mu$m due to larger bending loss and scattering loss occurring at the sidewalls at larger bend curvature. 

In Fig.~\ref{fig:QoverV}, we plot the cooperativity parameter $C$ as a result of the scan, assuming an atom is trapped at $(\rho_a, z_a) = (R,100)$~nm. With the surface roughness at its current value, as in fig. 3(a), $C$ is maximized when the waveguide geometry tends towards larger cross section ($W, H$), which reduces surface scattering, and smaller radius $R$, which reduces the mode volume. The best projection $C\approx 46$ uses a smaller ring $R \sim 10~\mu$m than in our current design and $(W,H)=(0.8,0.36)~\mu$m. For reduced surface roughness, as in Fig.~\ref{fig:QoverV}(b), a resonator geometry of $(W,H,R)=(1.1,0.29,16)~\mu$m, as shown in Fig.~\ref{fig1}, can achieve $Q \approx 4.5 \times 10^6$. The projected cooperativity reaches $C\approx 536$, an almost 12 times improvement from our current optimal value. Similarly, a fundamental transverse-electric (TE) mode can also be optimized with a different geometry, giving higher $Q \approx 8\times 10^6$ but with a lower optimal $C\approx 397$ due to larger $V_m$.

\section{Atom trapping in the optical nearfield of the microring platform}
\label{trap}
We now discuss two schemes, both capable of creating tight far off-resonant optical traps for cold atoms around $z_\mathrm{t}=100~$nm above the top surface of the microring. While either scheme can function fully independently, we discuss the combination of both schemes for atom array assembly on a microring (racetrack) resonator. 

\begin{figure*}[t]
\centering
\includegraphics[width=1.8\columnwidth]{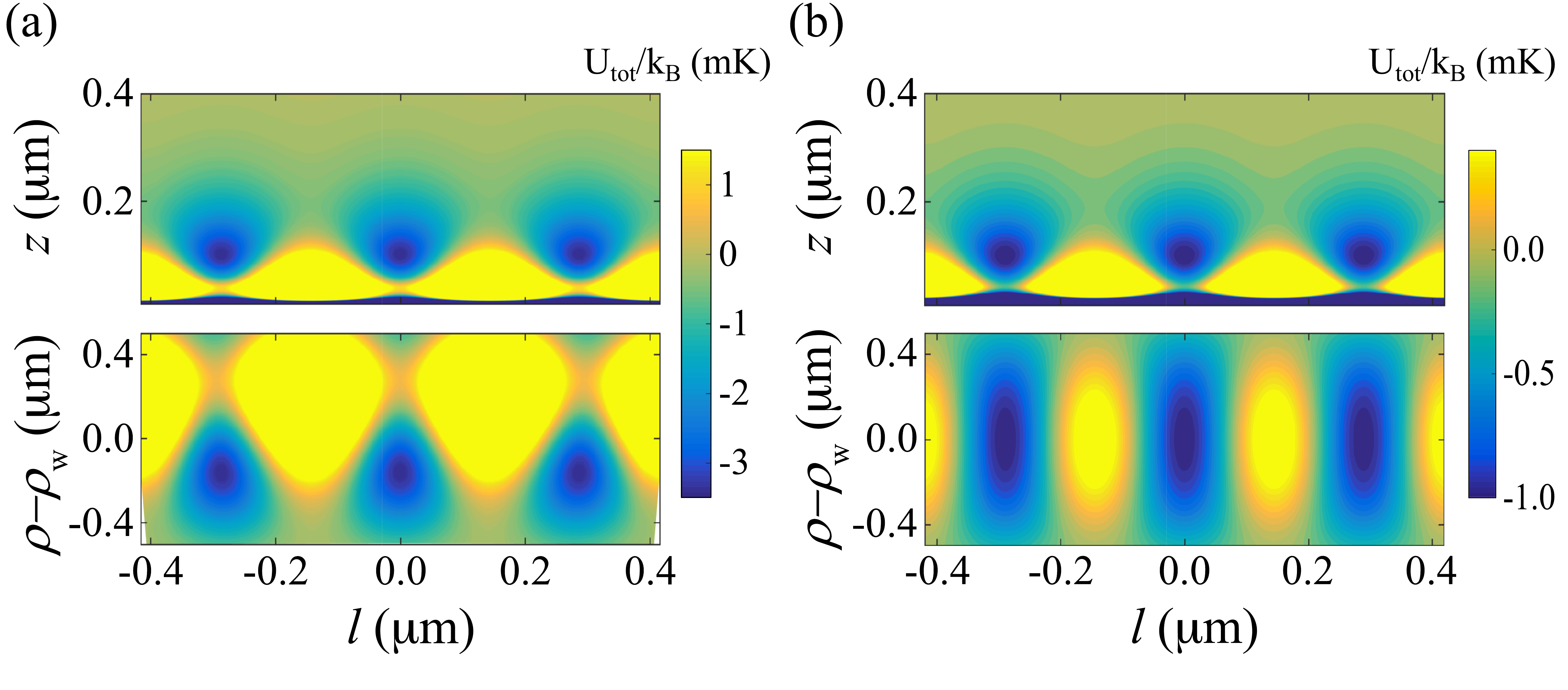}
\caption{Evanescent field lattice potential on (a) a microring and (b) a racetrack resonator, respectively. Top panels show the potential cross sections $U_\mathrm{tot}(\rho_\mathrm{t},l,z)$ while the bottom panels show $U_\mathrm{tot}(\rho,l,z_\mathrm{t})$, respectively. The lattice constant is $d = 290~$nm. Parameters used are identical to those of Fig.~\ref{fig:eft_xz}(a, c).}
\label{fig:lattice}
\end{figure*}

\subsection{State-insensitive two-color evanescent field trap}
The evanescent field trapping scheme shares similarities with those realized in nano-fiber traps \cite{le_kien_atom_2004,lacroute2012state}, and proposed in nanophotonic waveguides \cite{meng2015nanowaveguide,stievater2016modal}. The trap is formed by two TM modes excited near the `magic' wavelengths $\lambda_\mathrm{r} = 935.3$~nm and $\lambda_\mathrm{b} = 793.5$~nm, so that they do not create differential light shifts in the laser cooling transition of cesium (Supplement 1 Sec.~2). Here, $\lambda_\mathrm{b}$ (frequency $\omega_\mathrm{b}$) is blue-detuned from major optical transitions in the ground state, creating strongly repulsive optical force within a short range near the dielectric surface. $\lambda_\mathrm{r}$ (frequency $\omega_\mathrm{r}$) is red-detuned, leading to an attractive force with longer decay length than that of the $\lambda_\mathrm{b}$-mode. The combination of both modes creates a stable trap above the waveguide surface; see Fig.~\ref{fig:eft_xz}.

Along the microring, coherent back-scattering mixes the CW and CCW counter-propagating modes and converts an otherwise smooth evanescent field intensity profile into a standing wave pattern just like an optical lattice (Supplement 1 Sec.~1-B). An optical lattice potential can provide strong longitudinal trap confinement along the microring. Exciting the resonator from either end of the coupling waveguide bus with power $P_\mathrm{w}$ and frequency $\omega_\mathrm{b(r)}$ near a resonance $\omega_\mathrm{0,b(r)}$ creates an electric field with a corrugated intensity profile
\begin{equation}
|\mathbf{E}(\mathbf{r})|^2 = 
\mathcal{I}|\mathcal{E}(\rho,z)|^2\left[1 \pm V\sin(2 k l \pm \xi)\right],\label{eq:elattice}   
\end{equation}
where the $\pm$ sign is given by the direction of bus waveguide coupling that excites opposite mixtures of the resonator modes (Supplement 1 Sec. 1-D); the sign flip is necessary due to coherent back-scattering. Here $\mathcal{I}=\mathcal{I}(\alpha)$ is a near-resonance energy build-up factor, with $\alpha = \frac{\kappa}{2}+i\left[\omega-\omega_\mathrm{0,b(r)}\right]$ and a back-scattering rate $\beta$,
\begin{equation}
\mathcal{I}(\alpha)=\frac{\kappa_\mathrm{c}P_\mathrm{w}}{\hbar \omega}\frac{|\alpha|^2+\beta^2}{|\alpha^2+\beta^2|^2},\label{eq:int}
\end{equation}
$|\mathcal{E}(\rho,z)|$ is the normalized mode field amplitude, giving $2\epsilon_0 \int\epsilon(\mathbf{r})|\mathbf{E}(\mathbf{r})|^2 d\mathbf{r} = \hbar \omega \mathcal{I}$, $k$ is the propagation wavenumber, $l$ is the arc length along the microring waveguide, and $\xi= \tan^{-1}2(\omega-\omega_\mathrm{0,b(r)})/\kappa$ is a frequency dependent phase shift. The visibility of the corrugation is given by
\begin{equation}
V(\alpha)=2v\frac{|\alpha \beta|}{(|\alpha|^2+\beta^2)},\label{eq:vis}
\end{equation}
where the amplitude factor $v\approx 0.2$ for a TM-mode (Supplement 1 Sec.~1-B). For the simplicity of discussions, we assume the waveguide parameters are equal for the two color modes.

To form a homogeneous lattice trap along the resonator, we eliminate the standing wave pattern in the $\omega_\mathrm{b}$ mode to avoid incommensurate alignment between the blue-repulsive node and the red-attractive anti-node in the lattice potential. As shown in Fig.~\ref{fig:eft_xz}, we couple blue-detuned light from either end (+ and $-$) of the waveguide bus with symmetric detuning about $\omega_\mathrm{0,b}$ to completely cancel the potential corrugation (Eq.~\ref{eq:elattice}) \cite{cite:hlattice}. We note that a large detuning between the $+/-$ modes is necessary to eliminate their interference contribution to the trap potential.

We calculate the two-color evanescent field trap potential using the incoherent sum of two-color potentials as
\begin{align}
    U_\mathrm{ev}(\mathbf{r}) = & - \alpha^{(0)}_\mathrm{r}\mathcal{I}_\mathrm{r}|\mathcal{E}_\mathrm{r}(\rho,z)|^2\left[1+V\cos(2k_\mathrm{r} l)\right] \nonumber \\ 
    & - \alpha^{(0)}_\mathrm{b}\mathcal{I}_\mathrm{b}|\mathcal{E}_\mathrm{b}(\rho,z)|^2, \label{eq:Uev}
\end{align}
where $\alpha^{(0)}_\mathrm{r} \approx 3033~$(a.u.; in atomic unit) and $\alpha^{(0)}_\mathrm{b} \approx -2111~$ (a.u.) are atomic dynamic scalar polarizabilities at frequencies $\omega_\mathrm{r}$ and $\omega_\mathrm{b}$, respectively. Similarly, $\mathcal{E}_{r,(b)}$ are the excited mode fields, $\mathcal{I}_\mathrm{r}=\mathcal{I}^+_\mathrm{r}$ and $\mathcal{I}_\mathrm{b}=\mathcal{I}^+_\mathrm{b} + \mathcal{I}^-_\mathrm{b}$ are energy build-up factors of the two color modes (see Fig.~\ref{fig:eft_xz} (e, f)); $k_\mathrm{r}$ is the wave number of the red mode, and $l=0$ is shifted to center on a lattice site. 

The two-color evanescent field trap can be made state-insensitive, that is, independent of the Zeeman sublevels of cesium ground state atoms. We note that the vector light shift is completely canceled in the presented coupling scheme \cite{lacroute2012state}, as discussed in Supplement 1 Sec.~2-C. Thus, we only include the scalar light shift in Eq.~(\ref{eq:Uev}).

Figure~\ref{fig:eft_xz}(a) shows sample potential cross sections in a transverse plane above the microring. Due to finite curvature of the microring waveguide that results in non-equal center shifts of the two color modes, the trap center is shifted inwards by $\sim 160~$nm and the trap axes are rotated. To avoid trap distortion, a racetrack resonator design [Fig.~\ref{fig:eft_xz}(c)] can be employed, where a symmetric trap can be found above the linear segments of the racetrack. The trap centers in Fig.~\ref{fig:eft_xz}(a,c) are $(\rho_\mathrm{t}-\rho_\mathrm{w},z_\mathrm{t})\approx(-160,100)~$nm on a microring and $(\rho_\mathrm{t}-\rho_\mathrm{w},z_\mathrm{t})\approx(0,100)~$nm on a racetrack, respectively, where $(\rho,z)=(\rho_\mathrm{w},0)$ is the top surface center of the microring (racetrack) waveguide.

To illustrate that the trap is strong enough against the atom-surface attraction, we have included in Fig~\ref{fig:eft_xz} the contribution of a Casimir-Polder potential $U_\mathrm{cp}(\mathbf{r}) \approx -C_4/z^3(z + \lambdabar)$ for $|\rho-\rho_\mathrm{w}|\leq W/2$, where $C_4/h = 267~$Hz$\cdot\mu$m$^4$ is for cesium atom-\sitnf~surface coefficient and $\lambdabar=136~$nm is an effective wavelength \cite{Stern_2011}. The total trap potential
\begin{equation}
    U_\mathrm{tot}(\mathbf{r}) = U_\mathrm{ev}(\mathbf{r}) + U_\mathrm{cp}(\mathbf{r})
\end{equation}
is dominated by $U_\mathrm{cp}$ only when $z\lesssim50~$nm. Here, the trap opens at potential saddle points near $(\rho_\mathrm{s},z_\mathrm{s}) \approx  (-300,60)$~nm  for a ring and $(\pm 275,67)$~nm for a racetrack; $U_\mathrm{tot}(\rho_\mathrm{s},0,z_\mathrm{s})-U_\mathrm{tot}(\rho_\mathrm{t},0,z_\mathrm{t})$ defines the trap depth $\Delta U$, which is $\Delta U\approx k_\mathrm{B}\times 130~\mu$K in Fig.~\ref{fig:eft_xz}(a,c), and is 10 times larger than the typical temperature of laser-cooled cesium atoms.

The energy build-up factors used to calculate the trap on the microring (racetrack) in Fig.~\ref{fig:eft_xz}(a,c) are $\mathcal{I}^+_\mathrm{r}= 2.4 \times 10^5$ ($4.2\times 10^4$) for the $\omega_\mathrm{r}$ mode and $\mathcal{I}^+_\mathrm{b}=\mathcal{I}^-_\mathrm{b}= 2.7\times 10^5$ ($4.0\times 10^4$) for the $\omega_\mathrm{b}$ modes, respectively. Using the coupling scheme and parameters associated with Fig.~\ref{fig:eft_xz}(e,f), the required total power is $P_\mathrm{r} = 320~\mu$W (56~$\mu$W) for $\omega_\mathrm{r}$ and $P_\mathrm{b} = 740~\mu$W ($110~\mu$W) for $\omega_\mathrm{b}$ modes in a microring (racetrack), respectively.

 We note that by adjusting the power ratio of the two color modes, $z_\mathrm{t}$ can be moved away from or pulled closer to the waveguide surface. In Fig.~\ref{fig:eft_xz}(b,d), we keep $P_\mathrm{b}$ fixed while tuning the ratio $\mathcal{I}_\mathrm{r}/\mathcal{I}_\mathrm{b}$ (or $P_\mathrm{r}/P_\mathrm{b}$) and show that the trap center can be tuned from $z_\mathrm{t}\sim300~$nm to $<100~$nm. Meanwhile, $\rho_\mathrm{t}$ remains fairly unchanged. This important feature would allow us to initiate atom trapping and sorting at $z_\mathrm{t}>200~$nm and perform atom-light coupling at $z_\mathrm{t}<100~$nm, discussed later.

Figure~\ref{fig:lattice} shows the lattice potential along the axial position $l$ of a microring and a racetrack, plotted using the cross sections of $U_\mathrm{tot}$ in the planes of $\rho=\rho_\mathrm{t}$ and $z=z_\mathrm{t}$, respectively. The low visibility $V<1$ (Fig.~\ref{fig:eft_xz}(e)) in the attractive TM-mode keeps the lattice potential nearly attractive everywhere along the resonator until very close to the resonator waveguide surface $z<200~$nm. This feature allows atoms to traverse freely along the resonator without seeing strong potential barrier until they are cooled into individual lattice sites at $z_\mathrm{t}\approx 100~$nm. 

Overall, the evanescent field trap provides three-dimensional tight confinement. For the example given in Figs.~\ref{fig:eft_xz} and \ref{fig:lattice}, the trap frequencies are $(\omega_{\rho'}, \omega_l, \omega_{z'}) = 2\pi \times (175, 1500, 1180)~$kHz for a microring trap, where $\hat{\rho}'$ and $\hat{z}'$ are along the tilted axes due to trap distortion, and $(\omega_{\rho}, \omega_l, \omega_{z}) = 2\pi \times (80, 786, 681)~$kHz for a racetrack trap.

\subsection{Top-illuminating optical potential}
\begin{figure}[b!]
\centering
\includegraphics[width=0.8\columnwidth]{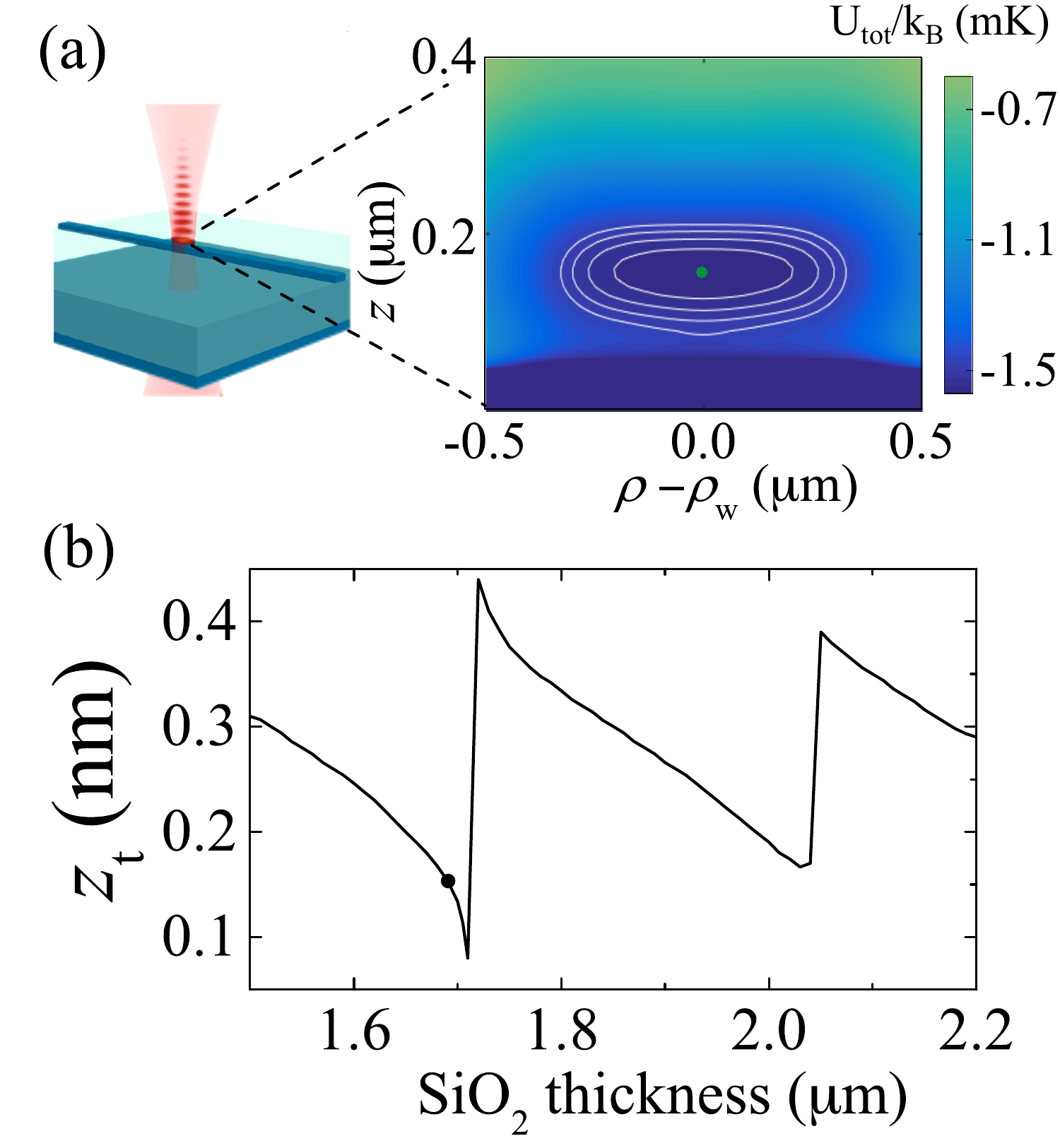}
\caption{Top-illuminating optical trap. (a) A tightly focused optical beam (an optical tweezer)  creates a lattice of microtraps on top of the resonator waveguide. Inset shows the total potential cross-section $U_\mathrm{tot}(\rho, z)$ of the nearest trap site above the surface. Green sphere marks the trap center at $z_\mathrm{t}=150~$nm. Potential contours are $K_B \times25$, 50, 75 and 100~$\mu$K above $U_\mathrm{tot}(\rho_\mathrm{w},z_\mathrm{t}) \approx  -k_\mathrm{B}\times 1.57 $~mK at the trap center. (b) Scanning the trap center $z_\mathrm{t}$ by tuning the thickness of the dioxide layer and keeping the thickness of the bottom nitride layer fixed at $600~$nm. Filled circle marks the geometry parameter for (a).}
\label{fig:reflection}
\end{figure}

Illuminating the microring from the top surface using a red-detuned beam (wavelength $\lambda_\mathrm{r}$) can also create a tight optical potential due to the top-illuminating beam interfering with its reflection from the microring structure (Fig.~\ref{fig:reflection}). The trap site closest to the dielectric surface, typically within a distance $<\lambda_\mathrm{r}/4$, can be utilized for trapping atoms in the near-field region of the resonator mode. Once an atom is trapped, the top-illuminating beam can also be steered in the horizontal plane to transport and organize atoms along the microring. This simple scheme need not have trapping light guided by the resonator, and can be universally applied to any dielectric structures with finite surface reflectance. The strength and position of the first trap site can in principle be finely adjusted through geometrically tuning the phase shift of the reflected light. In fact, this method has been successfully implemented in a number of pioneering experiments trapping atoms on suspended nanostructures \cite{thompson_coupling_2013, goban_superradiance_2015}, although fully independent trap tuning cannot be achieved because the geometry of a nanostructure needs to be adjusted and its desired guided mode property is inevitably affected.
\begin{figure*}[t!]
\centering
\includegraphics[width=1.5\columnwidth]{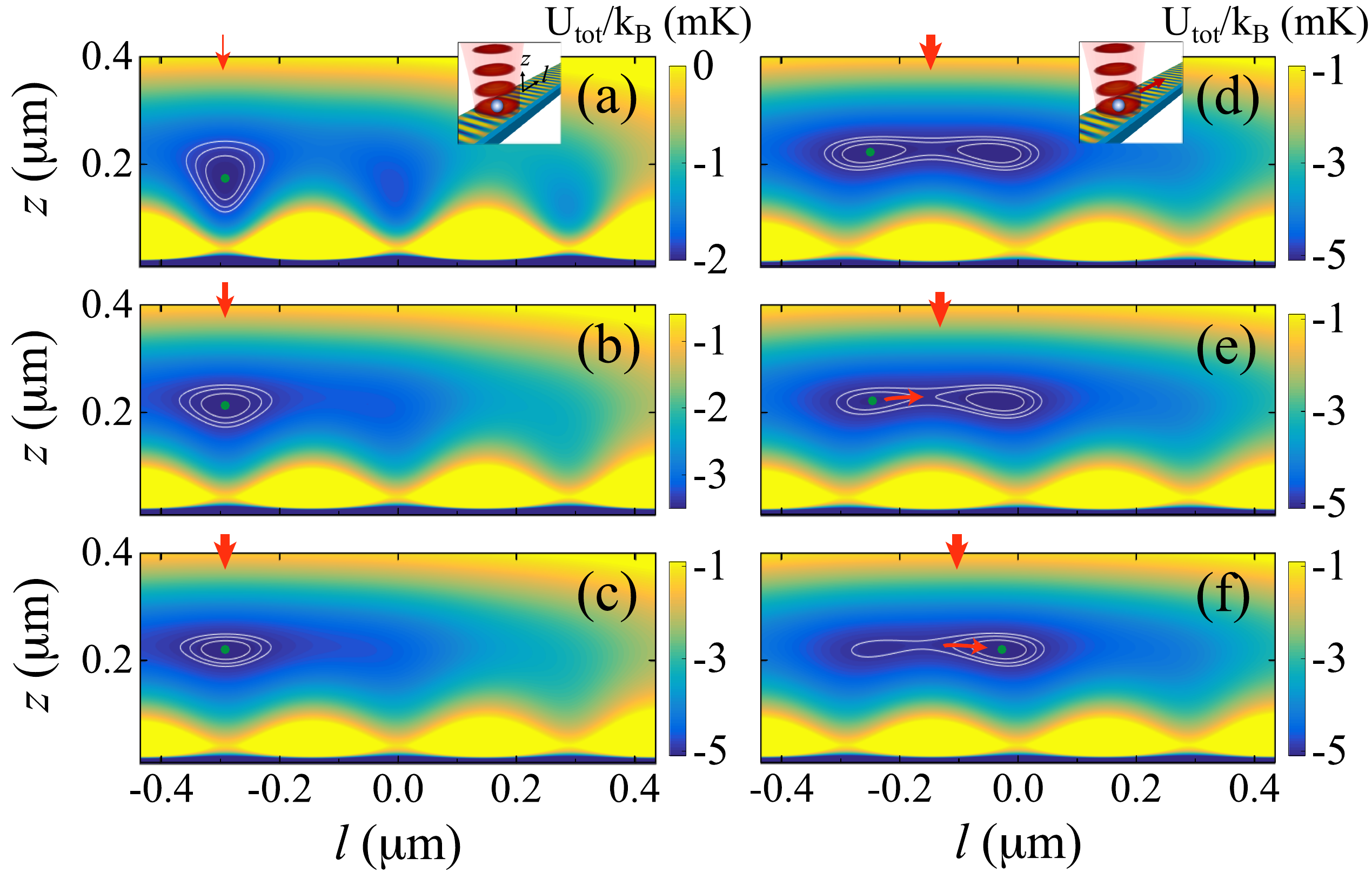}
\caption{Atom transport in an evanescent field lattice trap. (a-c) Illustration of tweezer trap transfer. Potential cross-sections $U_\mathrm{tot} = U_\mathrm{ev} + U_\mathrm{tw}+U_\mathrm{cp}$ are shown with a tweezer trap centered at $l=-d=-290~$nm and with increasing tweezer power $P_\mathrm{tw}=$ (a) 2~mW, (b) 4~mW, and (c) 6~mW, respectively.  (d-f) Trapped atom transport from $l=-d$ to $l=0$. Potential cross-sections $U_\mathrm{tot}$ are shown with a fixed tweezer power $P_\mathrm{tw}=6~$mW and shifted tweezer trap center at $l/d=$ (d) -0.5, (e) -0.44, and (f) -0.37, respectively. Vertical arrows indicate the center of the tweezer trap. Filled circles mark the position of trapped atoms in the target site center, which is enclosed by potential contours that are $K_B \times25$, 50, and 100~$\mu$K, respectively, above the local potential minimum at $z'_\mathrm{t}\approx 200~$nm.}
\label{fig:transport}
\end{figure*}
For the microring (racetrack) platform, trap condition in a top-illuminating potential can be finely adjusted independent of the waveguide properties, since multiple interfaces exist in the underlying membrane substrate. A desired trap condition can be realized simply by tuning the thickness of the dioxide or nitride layers in the membrane, as illustrated in Fig.~\ref{fig:reflection}.

Figure~\ref{fig:reflection}(a) shows a sample potential cross-section $U_\mathrm{tot}(\rho,z) = U_\mathrm{tw}(\rho,z) + U_\mathrm{cp}(\rho,z)$, where the optical potential $U_\mathrm{tw}$ is calculated by using a FDTD method \cite{cite:lumerical} with a tightly focused Gaussian beam ($\lambda_\mathrm{r}=935~$nm) of a $1/e^2$ beam waist $w=1.2~\mu$m and a power of $P=3.5~$mW projected from the top of the microring waveguide. The beam is polarized along $\hat{\rho}$, which is perpendicular to local waveguide orientation, to minimize reflection from the surface of the microring. In Fig.~\ref{fig:reflection}(b), we scan the thickness of the membrane, and illustrate a configuration such that the closest trap site to the microring surface is centered around $(\rho_\mathrm{t}-\rho_\mathrm{w},z_\mathrm{t}) \approx (0,150)~$nm, where $z_\mathrm{t}$ is significantly smaller than $\lambda_\mathrm{r}/4\approx234~$nm, and the trap depth of $\Delta U\approx k_\mathrm{B} \times 105~\mu$K. With a tightly focused beam waist, the top-illuminating beam forms a tweezer-like optical potential, providing also strong transverse (along $\hat{\rho}$) and axial (along $\hat{l}$) confinements. The former is due to the waveguide width ($W\sim \lambda_\mathrm{r}$), leading to a strong intensity variation in the transverse direction. The axial confinement, on the other hand, is ensured by the small beam waist of the tweezer beam. For the example given in Fig.~\ref{fig:reflection}(a), the trap frequencies are $(\omega_{\rho}, \omega_l, \omega_{z}) = 2\pi \times (43, 89, 332)~$kHz.

\subsection{Trap loading and atom sorting along a microring (racetrack) resonator}
In \cite{kim2019}, we have experimentally demonstrated that cold atoms can be directly laser-cooled on a membrane optical circuit and loaded into a top-illuminating optical tweezer trap. We note that the presence of lattice potential along a tweezer trap likely reduces the probability for cold atoms to be cooled directly into the first site near the microring surface. Instead, multiple atoms may be randomly confined along the lattice of microtraps within a tweezer. An optical conveyor belt can be implemented to transport trapped atoms onto the microring surface \cite{kim2019}. By monitoring the transmission of a resonator mode tuned to atomic resonance, it is possible to transport trapped atoms onto the micro-ring surface with deterministic control. 

On the other hand, a two-color evanescent field trap provides a smooth transverse potential landscape (along $\hat{z}$), allowing a large number of laser-cooled atoms to be loaded uninterruptedly from freespace into the lattice potential at $z_\mathrm{t}\approx 100~$nm above the microring, which has recently been demonstrated in nanofiber traps \cite{le_kien_atom_2004,balykin2004atom,lacroute2012state}. Nonetheless, these trapped atoms should randomly fill the optical lattice without organization.

In Fig.~\ref{fig:transport}, we illustrate how a tweezer trap can be used to sort trapped atoms in an evanescent field trap, similar to those in an optical lattice in freespace \cite{barredo_atom-by-atom_2016}. To begin with, one may utilize the two-color evanescent field trap for initial atom loading into $U_\mathrm{tot}= U_\mathrm{ev}+U_\mathrm{cp}$ at $z_\mathrm{t}\approx 100\sim 200~$nm. Following laser cooling, fluorescence imaging \cite{kim2019} can be performed to determine the atomic distribution along the resonator. Once identifying the location of all trapped atoms, an optical tweezer trap $U_\mathrm{tw}$ can be ramped on to draw an atom into a new vertical position $z'_\mathrm{t}\approx 200~$nm (Fig.~\ref{fig:transport}(a-c)), and transport it along the resonator into a designated lattice site (Fig.~\ref{fig:transport}(d-f); across multiple sites). Following transport, the tweezer beam can then be adiabatically ramped off, releasing the trapped atom back to the evanescent field trap at $z_\mathrm{t}$. Atom-sorting can be realized by reiterating the procedures to reorganize atoms in different trap sites.

\section{Conclusion and outlook}
In this paper, we have demonstrated that microring and racetrack resonator platforms can be fabricated to be completely compatible with laser cooling and trapping with cold atoms and with reasonably high cooperativity parameters $C\approx 25 \sim 46$. This number can be further boosted by more than 10-fold with further fabrication improvements, thus holding great promises as an on-chip atom cavity QED platform. We have discussed two viable optical trapping schemes, both using magic wavelengths of atomic cesium, for localizing atoms around $z_\mathrm{t}=75\sim 150~$nm above the dielectric surface of a resonator waveguide structure. The combination of both schemes permits controlled atom transport along a resonator, allowing for the formation of an organized atom-nanophotonic hybrid lattice useful for collective quantum optics and many-body physics \cite{ChangRMP2018,hung_quantum_2016}. 

Lastly, we note that although our emphasis is on coupling with cold trapped atoms, these microrings may also be adapted for  coupling with solid state quantum emitters \cite{wei2015silicon,saskin2019narrow,nandi2019anomalous}, or with atomic thermal vapors \cite{ritter2016coupling,ritter2018coupling}. For emitters on the surface of a resonator waveguide and considering only the radiative losses, the effective mode volume $V_\mathrm{m} \approx 63~(\lambda/n)^3$ and $C \approx 360~n^{-3}$ using our current fabricated structures, where n is the host refractive index for embedded quantum emitters. Improving to $Q \approx 4.5 \times 10^6$ would lead to a  projected $C \gtrsim 5000~n^{-3}$ that may be potentially useful for on-chip solid-state quantum photonics.

\section*{Acknowledgements}
We acknowledge discussions from H. J. Kimble, S.-P. Yu, S. Bhave, M. Hosseini, S. Caliga, Y. Xuan, and B.-L. Yu. Funding is provided by the AFOSR YIP (Grant NO. FA9550-17-1-0298), ONR (Grant NO. N00014-17-1-2289) and the Kirk Endowment Exploratory Research Recharge Grant from the Birck Nanotechnology Center.

\bibliography{fab}

\include{Appendix}

\end{document}

%% file: Appendix.tex
\renewcommand\thefigure{\thesection.\arabic{figure}}    
\setcounter{figure}{0} 

\appendix
\section{Electric field profile and mode-mixing in a microring resonator}\label{appendix:profile_and_mixing}
\subsection{Mode profile in an ideal microring}
\label{appendix:mode}
Due to the small dimensions of our microring geometry, it supports only fundamental modes within the resonator waveguide at the wavelengths of our interest. A perfect microring supports resonator modes of integer azimuthal mode number $m$, whose electric field can be written as $\mathbf{E}_\pm(\mathbf{r},t) = \mathbf{E}_\pm(\mathbf{r}) e^{-i\omega t}$, where the spatial field profile in cylindrical coordinates is
\begin{equation}
    \mathbf{E}_\pm(\mathbf{r}) = \left[ \mathcal{E}_\rho (\rho,z)\hat{\rho} \pm i \mathcal{E}_\phi(\rho,z)\hat{\phi} + \mathcal{E}_z (\rho,z)\hat{z}\right]e^{\pm i m\phi}.\label{appendix:mode:A1}
\end{equation}
We additionally require that the mode field satisfies the normalization condition $2\epsilon_0\int \epsilon(\mathbf{r})|\mathbf{E}_\pm(\mathbf{r})|^2d\mathbf{r}= \hbar\omega$, where $\epsilon_0$ is the vacuum permittivity and $\epsilon(\mathbf{r})$ is the dielectric function. Here, $\mathcal{E}_\mu(\rho,z)$ ($\mu=\rho,\phi,z$) are real functions and are independent of $\phi$ due to cylindrical symmetry. We note that the azimuthal field component is $\pm\pi/2$ out of phase with respect to the transverse fields due to strong evanescence field decay and transversality of the Maxwell's equation. The perfect resonator modes are traveling waves and the $\pm$ sign indicates the direction of circulation. The mode fields of opposite circulations are complex conjugates of one another $\mathbf{E}_+(\mathbf{r})=\mathbf{E}^*_-(\mathbf{r})$. We can assign a propagation number $k = m\phi /l$, where $l= R\phi$ is the arc length.
 
\begin{figure}[b]
\centering
\includegraphics[width=1\columnwidth]{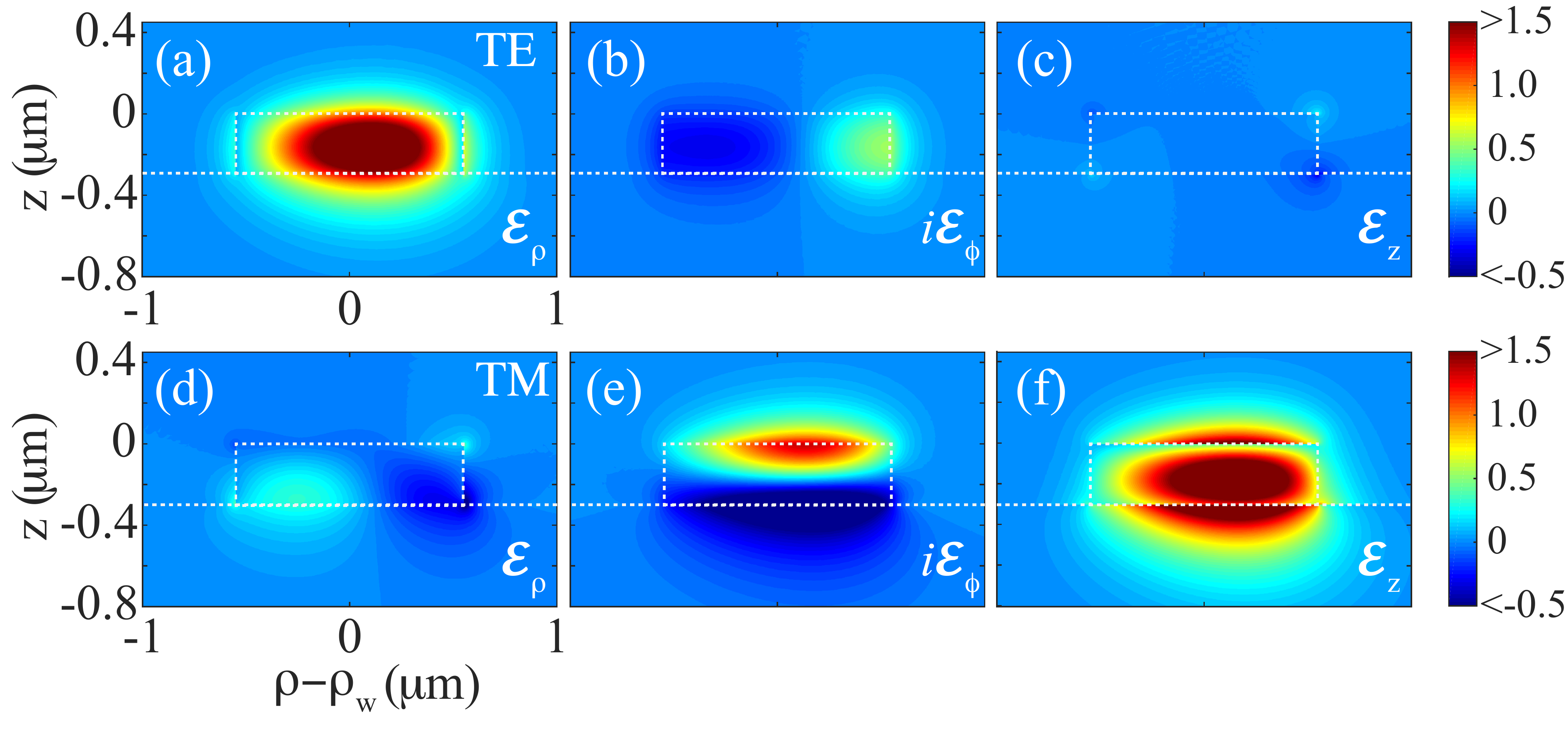}
\caption{Electric field vector components $(\mathcal{E}_\rho,i\mathcal{E}_\phi,\mathcal{E}_z)$ of (a-c) the fundamental transverse electric (TE) mode and (d-f) the transverse magnetic (TM) mode (in arbitrary units). Dashed lines mark the boundaries of the dielectrics. Radius of the ring $R=\rho_\mathrm{w}=16~\mu$m.}
\label{fig:TETMEi}
\end{figure}

The electric field functions are evaluated using a software employing a finite element method (COMSOL) \cite{4230891,cheema2010implementation}. Fig.~\ref{fig:TETMEi} shows the electric field components (in cylindrical coordinates) of the fundamental transverse electric (TE) and transverse magnetic (TM) resonator modes. The fields are slightly asymmetric across the center of the waveguide at $\rho = \rho_\mathrm{w}$ due to finite curvature of the microring (radius $R=16~\mu$m). We also note that the out-of-phase axial component $i\mathcal{E}_\phi$ is stronger in the TM-mode, resulting from stronger evanescent field along the $\hat{z}-$axis where the waveguide confinement is strongly subwavelength.

\subsection{Mode mixing in a microring}
\label{appendix:ModeMixing}
In the presence of fabrication imperfections, surface scatters induce radiation loss and mode mixing. The former, together with other intrinsic loss mechanisms (discussed in \cite{Borselli:05,Ji:17,Pfeiffer:18}), induces intrinsic resonator energy loss at a rate $\kappa_\mathrm{i}$. The latter effect can be treated perturbatively, with photons scattered from one resonator mode into another.  Assuming small dielectric irregularities in a high-Q resonator, only counter-propagating modes with identical azimuthal number can couple via back-scattering from the surface roughness (at a rate $\beta$). In this paper we obtain this back-scattering rate experimentally. 

To understand mode-mixing and its impact on the resonator mode profiles, we apply well-established coupled mode theory \cite{Srinivasan2007} for two counter propagating modes of interest. Using $a_\mathrm{-}(a_\mathrm{+})$ to denote the amplitude of the clock-wise (CW) and counter clock-wise (CCW) propagating resonator modes in a mode-mixed resonator field 
\begin{equation}
\mathbf{E}(\mathbf{r},t) = a_+(t) \mathbf{E}_+ (\mathbf{r},t)+a_-(t) \mathbf{E}_- (\mathbf{r},t),   
\end{equation}
we have the following coupled rate equation
\begin{align}
     \frac{da_\mathrm{+}}{dt} = & - \left(\frac{\kappa}{2}  + i \Delta \omega\right) a_\mathrm{+} (t) + i \beta e^{i\xi} a_\mathrm{-}(t)\nonumber \\
     \frac{da_\mathrm{-}}{dt} = & - \left(\frac{\kappa}{2} + i \Delta \omega\right) a_\mathrm{-} (t) + i \beta e^{-i\xi} a_\mathrm{+}(t),\label{eqSM:couple}
\end{align}
where $\Delta\omega=\omega-\omega_0$ is the frequency detuning from the bare resonance $\omega_0$, $\beta$ is the coherent back-scattering rate, and $\xi$ is a scattering phase shift. The total loss rate $\kappa = \kappa_\mathrm{i} + \kappa_\mathrm{c}$ includes resonator intrinsic loss rate $\kappa_\mathrm{i}$ and the loss rate $\kappa_\mathrm{c}$ from coupling to the bus waveguide.

Due to the back-scattering terms in Eq.~\ref{eqSM:couple} mixing CW and CCW modes, a new set of normal modes are established whose rate equations are decoupled from each other and the frequencies of the new modes are shifted by $\beta$ and $-\beta$ relative to the unperturbed resonance, respectively. The electric field of the mixed mode can be written as $\mathbf{E}_{i}(\mathbf{r},t) = \mathbf{E}_{i}(\mathbf{r})e^{-i\omega t}$ ($i=1, 2$), where
\begin{align}
    \mathbf{E}_1(\mathbf{r})
            = & \left[(\mathcal{E}_\rho \hat{\rho} + \mathcal{E}_z\hat{z})\cos(m\phi +\frac{\xi}{2}) - \mathcal{E}_\phi \sin(m\phi + \frac{\xi}{2})\hat{\phi}\right] \nonumber \\
    \mathbf{E}_2(\mathbf{r}) 
    = & i\left[(\mathcal{E}_\rho \hat{\rho} + \mathcal{E}_z\hat{z})\sin(m\phi +\frac{\xi}{2})+ \mathcal{E}_\phi \cos(m\phi + \frac{\xi}{2})\hat{\phi}\right],\label{eqSM:mixed}
\end{align}
and we have dropped an overall factor $\sqrt{2}e^{-i\frac{\xi}{2}}$ for convenience. The fields in Eq.~\ref{eqSM:mixed} should also satisfy the normalization condition. With the presence of back-scattering, the resonator mode polarization now becomes linear but is rotating primarily in the $\rho$-$\phi$ ($z$-$\phi$) plane for TE (TM) mode along the microring. 

\begin{figure}[t]
\centering
\includegraphics[width=1\columnwidth]{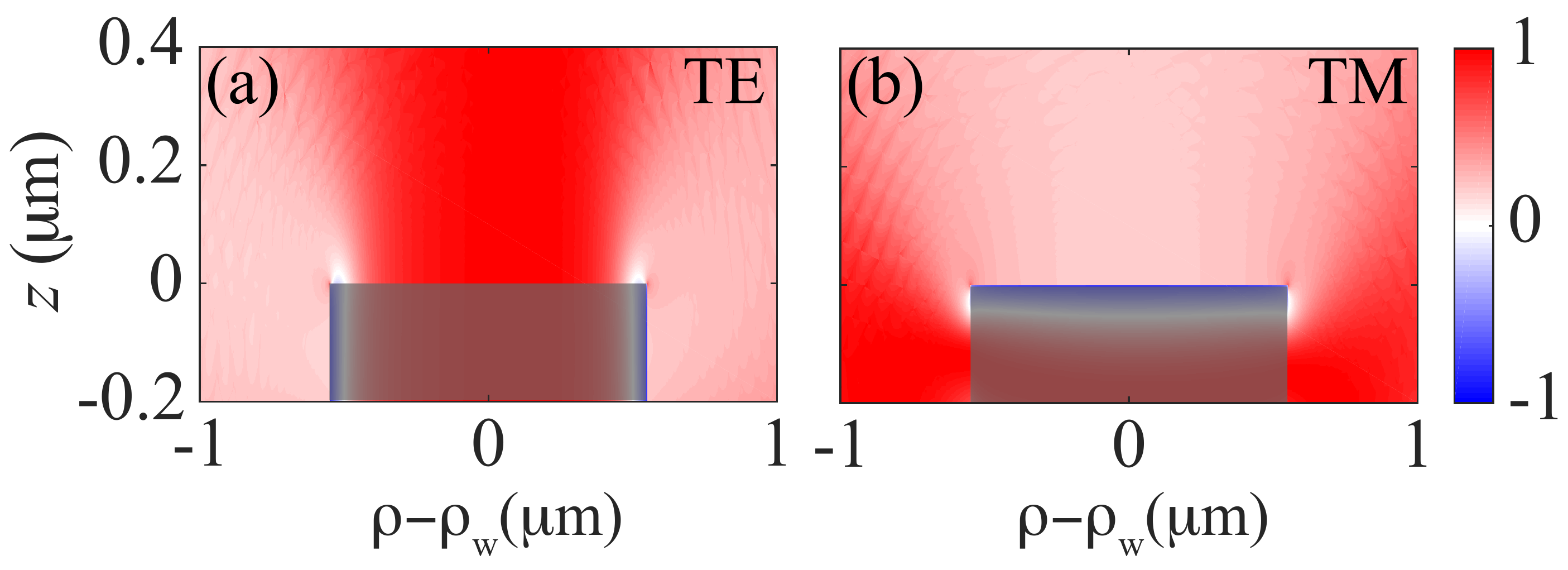}
\caption{The visibility amplitude factor $v(\rho,z)$ in the intensity profiles $\mathbf{E}_{1(2)}$ of the mixed TE (a) and TM (b) resonator modes, respectively. Shaded areas mark the microring waveguide.}
\label{fig:Vis}
\end{figure}

\subsection{Atom-photon coupling in a microring resonator}
\label{appendix:coupling}
We consider the atom-photon coupling strength
\begin{equation}
    g_i = d_i\sqrt{\frac{\omega}{2\hbar\epsilon_0V_\mathrm{m}}}\label{eqSM:g}
\end{equation}
where $d_i=\langle e|\mathbf{d}|g\rangle\cdot \mathbf{u}_i$ is the transition dipole moment, $\mathbf{u}_i=\mathbf{E}_i(\rho_\mathrm{a},z_\mathrm{a})/|\mathcal{E}(\mathbf{r}_a)|^2$ is the electric field polarization vector, $|\mathcal{E}|^2=|\mathcal{E}_\rho|^2+|\mathcal{E}_\phi|^2+|\mathcal{E}_z|^2$, $\hbar$ is the Planck constant divided by $2\pi$, and $V_\mathrm{m}(\mathbf{r}_a)$ is the effective mode volume at atomic position $\mathbf{r}_a$,
\begin{align}
    V_\mathrm{m}(\rho_\mathrm{a},z_\mathrm{a}) = & \frac{\int \epsilon(\rho,z)|\mathcal{E}(\rho,z)|^2 \rho d\rho dzd\phi}{\epsilon(\rho_\mathrm{a},z_\mathrm{a})|\mathcal{E}(\rho_\mathrm{a},z_\mathrm{a})|^2} \nonumber \\
     = & A_\mathrm{m}(\rho_\mathrm{a},z_\mathrm{a}) L.
\end{align}
Here $A_\mathrm{m}$ follows the definition Eq.~1 in the main text and $L=2\pi R$ is the circumference of the microring. 

We note that the coupled modes in Eq.~\ref{eqSM:g} can be the CW and CCW modes, that is $i=\pm$, when $g_i \gg \beta$. On the other hand, if $\beta \gg g_i$, an atom should be coupled to a mixed mode with $i=1, 2$. Our microring platform corresponds to the latter case. We also note that the exact value of the transition dipole moment $d_i$ depends on the atomic location and dipole orientation. In the main text, we simply replace $d_i$ with the reduced dipole moment $d$, where $d^2 = \frac{3\lambda^3\epsilon_0\hbar\gamma}{8\pi^2}$, and arrive at 
\begin{equation}
g = \sqrt{\frac{3\lambda^3\omega\gamma}{16\pi^2V_\mathrm{m}}}.    
\end{equation}

\subsection{Exciting the resonator mode via an external waveguide}
\label{appendix:exciting_mode}
If we consider exciting the resonator mode with input power $|s_\pm|^2$ (actual power $P_\mathrm{w}$ normalized with respect to $\hbar \omega$) from either end of the bus waveguide, additional amplitude growth rate $K s_\pm$ can be added to the right hand side of Eq.~\ref{eqSM:couple}; in the case of a lossless coupler $K=i\sqrt{\kappa_\mathrm{c}}$. Due to phase matching conditions between the linear waveguide and the microring, $s_+(s_-)$ only couples to the CCW(CW) mode and not to the other mode of opposite circulation. In the original CCW/CW basis, the mode amplitudes are
\begin{align}
     a_+ = & K\frac{\alpha s_+ + i \beta s_-}{\alpha^2 +\beta^2} \nonumber \\
     a_- = & K \frac{i\beta^* s_+ + \alpha s_- }{\alpha^2 +\beta^2},
\end{align}
where $\alpha = \frac{\kappa}{2}+i\Delta\omega $. If we now consider exciting the resonator modes from one side of the bus waveguide, the intra-resonator field is
\begin{equation}
     \mathbf{E}^\pm(\mathbf{r}) =  \frac{\alpha Ks_\pm}{\alpha^2+\beta^2}\left[ \mathbf{E}_\pm(\mathbf{r}) + \frac{i\beta e^{\mp i\xi}}{\alpha}\mathbf{E}_\mp(\mathbf{r})\right],\label{eqSM:extE}
\end{equation}
The $\pm$ sign in Eq.~(\ref{eqSM:extE}) indicates either $|s_+| > 0$ (and $s_-=0$) or $|s_-|>0$ (and $s_+ =0$). With back-scattering mixing counter-propagating modes, the field intensity is a standing wave
\begin{equation}
     |\mathbf{E}^\pm(\mathbf{r})|^2 = \mathcal{I}|\mathcal{E}(\rho,z)|^2 \left[ 1 \pm V(\rho,z)\sin(2m\phi+\xi_\pm)\right],
     \label{eqSM:extEsqr}
\end{equation}
where $\xi_\pm = \xi \pm \arg(\alpha)$. The sign flip in the intensity corrugation is due to the opposite mixtures of the resonator modes being excited, Eq. (\ref{eqSM:extE}), and an overall $\pi/2$ phase shift in the back-scattered mode. We have a frequency-dependent energy build-up factor
\begin{equation}
    \mathcal{I}= \mathcal{I}_0\frac{|\alpha|^2+\beta^2}{|\alpha^2+\beta^2|^2},
\end{equation}
where $\mathcal{I}_0 = \frac{\kappa_\mathrm{c} P_\mathrm{w}}{\hbar\omega}$ for a lossless coupler and
\begin{equation}
    V(\rho,z)=\frac{2|\alpha|\beta}{|\alpha|^2+\beta^2}v(\rho,z)\label{eqSM:vis}
\end{equation}
is the visibility of the standing wave; $V\leq v$ and equality holds when $|\alpha|=\beta$. Here, $v=1-2|\mathcal{E}_\phi|^2/|\mathcal{E}|^2$ is a visibility amplitude factor. The presence of the axial field reduces the visibility of the standing wave: $v$ vanishes when $|\mathcal{E}_\phi|^2=|\mathcal{E}|^2/2$ and is largest with $|\mathcal{E}_\phi|^2=0$. As shown in Fig.~\ref{fig:Vis}, the visibility of the TE mode is $v\approx 1$ above the microring due to the smallness of the axial component. On the contrary, for TM-mode a smaller $v\approx 0.2$ above the waveguide results from large $|\mathcal{E}_\phi|^2$ as seen in Fig.~\ref{fig:TETMEi}. 

From Eqs.~(\ref{eqSM:extEsqr}-\ref{eqSM:vis}), we can develop schemes to maximize or minimize standing-wave visibility for evanescent field trapping. This is discussed in the main text and in the following sections. 

\section{AC Stark shift in an evanescent field trap}
\label{appendix:lightshift}
\subsection{Scalar and vector light shifts in the ground state}
When a ground state atom is placed above a microring with a strong evanescent field that is far-off resonant from the atomic resonances, it experiences a spatially varying AC stark shift

\begin{equation}
    U(\mathbf{r}) = -\alpha_{\mu \nu}(\omega) E_\mu(\mathbf{r})  E^*_\nu (\mathbf{r}),\label{eqSM:SS}
\end{equation}
where $\alpha_{\mu\nu}$ is the dynamic polarizability tensor and $E_\mu = \mathbf{E}\cdot \hat{\mathbf{e}}_\mu$ is the vector components of the microring evanescent field. In the irreducible tensor representation, the above tensor product can be separated into contributions from scalar (rank-0), vector (rank-1), and tensor (rank-2) terms
\begin{equation}
U(\mathbf{r})= U^s(\mathbf{r}) + U^v(\mathbf{r}) + U^t(\mathbf{r}),
\end{equation}
where
\begin{align}
      U^s(\mathbf{r}) = & -\alpha^{(0)}(\omega)|\mathbf{E}(\mathbf{r})|^2 \\
      U^v(\mathbf{r}) = & -i\alpha^{(1)}(\omega)\frac{\mathbf{E}(\mathbf{r})\times \mathbf{E}^*(\mathbf{r})\cdot\hat{\mathbf{F}}}{2F} \\
      U^t(\mathbf{r}) = & -\alpha^{(2)}(\omega)\frac{3}{F(2F-1)} \times \nonumber \\
      & \left[\frac{\hat{F}_\mu \hat{F}_\nu + \hat{F}_\nu \hat{F}_\mu}{2} -\frac{\hat{\mathbf{F}}^2}{3}\delta_{\mu\nu}\right]E_\mu E^*_\nu,
\end{align}
and $\alpha^{(0,1,2)}(\omega)$ are the corresponding scalar, vector, and tensor polarizabilities, $\hat{\mathbf{F}}$ is the total angular momentum operator, and $F$ is the quantum number. We note that, for ground state atoms in the $S$ angular momentum state, $\alpha^{(2)}=0$. Therefore we do not consider $U^t$ throughout the discussions. The calculations of $\alpha^{(0,1)}$ follow those of \cite{ding2012corrections},  using transition data summarized within, 
and is not repeated here. Table~\ref{alphatable} lists the value of polarizabilities used in the trap calculation.

\begin{table}[h]
    \centering
    \begin{tabular}{c|c|c|c}
        $\lambda$ & $\alpha^{(0)}$ (a.u.) & $\alpha^{(1)}$ (a.u.) & $\alpha^{(1)}/\alpha^{(0)}$\\  \hline
         $\lambda_\mathrm{r} $ & 3033 & -1632  & -0.5382\\
         $\lambda_\mathrm{b} $ & -2111 & -643.0  & 0.3046\\ 
    \end{tabular}
    \caption{Cesium 6S$_{1/2}, F=4$ ground state dynamic polarizabilities at $\lambda_\mathrm{r} =935.3~$nm and $\lambda_\mathrm{b}=793.5~$nm.}
    \label{alphatable}
\end{table}

\subsection{Scalar and vector light shifts in an evanescent field trap}
To form an evanescent field trap, the microring must be excited through an external waveguide. Equations~(\ref{eqSM:extE}-\ref{eqSM:extEsqr}) can be used to calculate the single-end excited resonator electric field. The complex polarization of a mixed resonator mode induces both scalar and vector components of the AC Stark shift. Using $\mathbf{E}(\mathbf{r})=\mathbf{E}^\pm(\mathbf{r})$ from Eq.~(\ref{eqSM:extEsqr}), the scalar light shift forms a standing-wave potential 
\begin{align}
    U^s(\mathbf{r}) = -\alpha^{(0)}(\omega) 
    \mathcal{I}|\mathcal{E}(\rho,z)|^2\left[ 1 \pm V(\rho,z)\sin(2m\phi+\xi_\pm)\right].\label{eqSM:vsc}
\end{align}
Meanwhile, the vector light shift depends on the cross product between the CW and CCW components in the excited field
\begin{equation}
     \mathbf{E}^\pm(\mathbf{r})\times \mathbf{E}^\pm(\mathbf{r})^* =
     \mp2i\tilde{\mathcal{I}}\mathcal{E}_\phi(\rho,z)\left[\mathcal{E}_\rho(\rho,z)\hat{z}-\mathcal{E}_z(\rho,z)\hat{\rho}\right],
\end{equation}
which is smooth along the microring (independent of $\phi$ coordinate) and varies only in the transverse coordinates $(\rho,z)$. Here, $\tilde{\mathcal{I}}$ is the build up factor for the vector potential
\begin{equation}
     \tilde{\mathcal{I}}=\mathcal{I}_0\frac{|\alpha|^2-\beta^2}{|\alpha^2+\beta^2|^2}.\label{eqSM:tildeI}
\end{equation}
In a special case when the atomic principal axis lies along the $\hat{z}$-axis, the vector light shift can be explicitly written as 
\begin{align}
     U^v = & \mp\alpha^{(1)}(\omega)
     \tilde{\mathcal{I}}\mathcal{E}_\phi(\rho,z)\times \nonumber \\
     & \left[\mathcal{E}_\rho(\rho,z)\frac{\hat{F}_z}{F}-\mathcal{E}_z(\rho,z)\frac{(\hat{F}_++\hat{F}_-)}{2F}\right],\label{eqSM:vec}
\end{align}
where $\hat{F}_\pm$ are the angular momentum ladder operators. 

The explicit dependence on angular momentum operators in $U^v$ reveals a diagonal, state-dependent energy shift and off-diagonal coupling terms. Near the anti-nodes of a standing wave Eq.~(\ref{eqSM:extEsqr}), which should serve as trap centers, the ratio between the vector and the scalar light shifts is found to be (dropping $\hat{\mathbf{F}}$-related factors)
\begin{equation}
     \left|\frac{U^v_\mu}{U^s}\right| \sim \frac{\alpha^{(1)}(\omega)}{\alpha^{(0)}(\omega)} \times 
     \frac{\mathcal{E}_\phi(\rho,z)\mathcal{E}_\mu(\rho,z)}{2|\mathcal{E}(\rho,z)|^2} \frac{\tilde{\mathcal{I}}}{\mathcal{I}},\label{eqSM:alphaRatio}
\end{equation}
where $U^v_\mu~(\mu=\rho, z)$ represent the amplitudes of the diagonal and off-diagonal terms in the vector light shift Eq.~(\ref{eqSM:vec}), respectively. 

\begin{figure}[t]
\centering
\includegraphics[width=1\columnwidth]{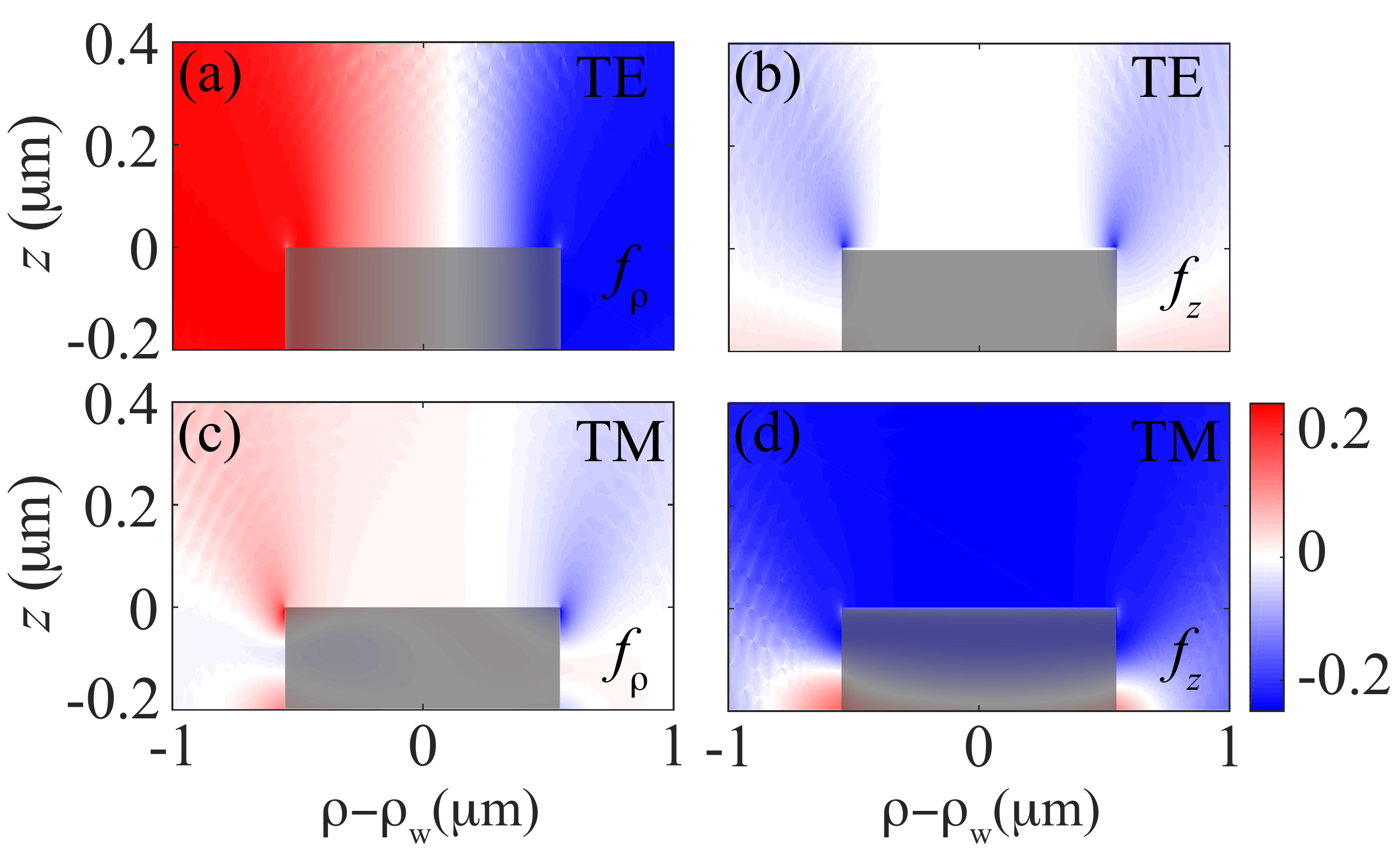}
\caption{Vector light shift polarization factor $f_\mu(\rho,z)$ for single-end excited (a-b) TE and (c-d) TM modes, respectively. Shaded areas mark the microring waveguide.}
\label{fig:Vec}
\end{figure}

Equation~(\ref{eqSM:alphaRatio}) suggests that the state dependent vector light shift can be smaller than the scalar shifts. For far-off-resonant light with frequency $\omega$ that is largely red- or blue-detuned from both cesium D1 and D2 lines, the vector polarizability $\alpha^{(1)}(\omega) < \alpha^{(0)}(\omega)$; see Table~\ref{alphatable}. The electric field polarization factor
\begin{equation}
f_\mu(\rho,z) = \frac{\mathcal{E}_\phi(\rho,z)\mathcal{E}_\mu(\rho,z)}{2|\mathcal{E}(\rho,z)|^2}     
\end{equation}
provides additional suppression. As shown in Fig.~\ref{fig:Vec}, a TE mode supports an off-diagonal factor $f_\rho \approx 0$ and the diagonal factor $f_z \approx 0.07$. For a TM mode, $f_\rho \approx -0.24$ and $f_z \approx 0.01$.

\subsection{Eliminating the vector light shifts}
\label{appendix:eliminating_vec}
\begin{figure}[t]
\centering
\includegraphics[width=1\columnwidth]{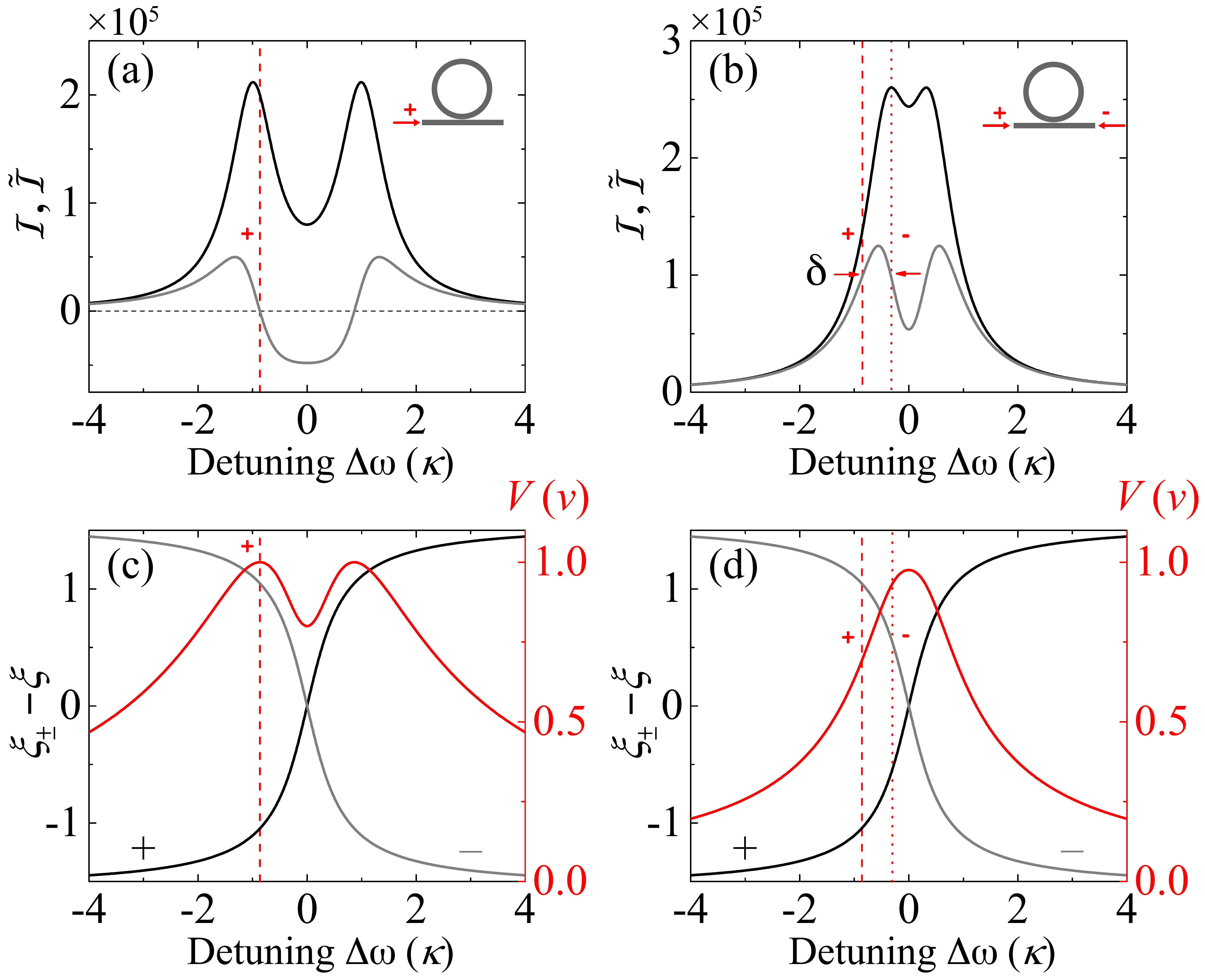}
\caption{Coupling schemes to eliminate the vector light shifts for microrings with $\beta > \kappa/2 $ (a,c) and $\beta < \kappa/2 $ (b,d).
(a,b) Intensity build-up factors $\mathcal{I}$ (black curve) and $\tilde{\mathcal{I}}$ (gray curve) for scalar and vector light shifts, respectively, for $\kappa/2\pi=1~$GHz, and $\beta/2\pi=$ (a)1~GHz and (b)0.4~GHz; $\mathcal{I}_0=10^5\kappa^2$. Vertical dash (+) and dotted (-) lines in red indicate the frequency detuning of excited modes for zeroing the vector shifts. $\pm$ signs denote the direction of excitation and the insets illustrate the corresponding coupling schemes. (c,d) Visibility $V$ (red curve) and phase shifts $\xi_\pm-\xi$ (black and gray curves) in the standing wave scalar potential Eq.~(\ref{eqSM:vsc}) under the same parameters and coupling schemes as in (a,b), respectively.}
\label{fig:int}
\end{figure}

In practical experiments, a state-independent trap is much preferred since it prevents parasitic effects such as dephasing or trap heating. To fully eliminate the vector shift, a straightforward method is to choose a proper detuning such that $\beta=|\alpha|$ (provided that $\beta>\kappa/2$) and $\tilde{\mathcal{I}}=0$, as suggested by Eqs.~(\ref{eqSM:tildeI}-\ref{eqSM:vec}) and illustrated in Fig.~\ref{fig:int} (a). Visibility $V$ is at the same time maximized as $|\alpha|=\beta$ creates equal superposition of CW and CCW modes up to a relative phase shift, as seen in Eq.~\ref{eqSM:extE}. The excited field becomes linearly polarized with spatially rotating polarization, similar to the form in Eq.~\ref{eqSM:mixed}, and leads to zero vector shift. In this simple scheme, the scalar light shift build-up factor is also near its maximal value, as in Fig.~\ref{fig:int}(a, c).

\begin{figure}[t]
\centering
\includegraphics[width=1\columnwidth]{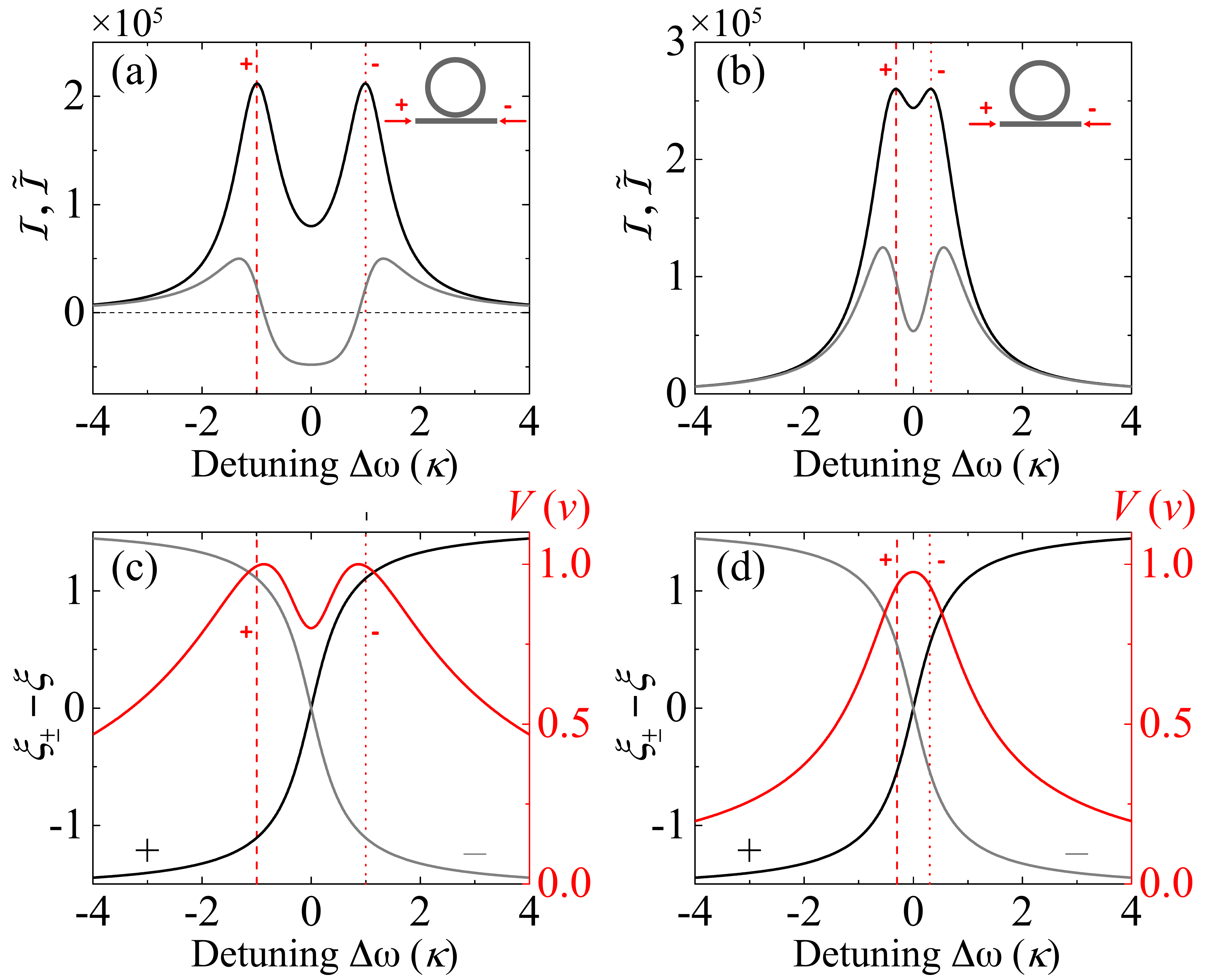}
\caption{Coupling schemes to eliminate both the vector light shifts and the standing wave pattern in the scalar potential. Same curves and parameters as those in Fig.~\ref{fig:int} are plotted, but with different coupling schemes (insets) and frequencies marked by the vertical dash (+) and dotted (-) lines.}
\label{fig:int_nolattice}
\end{figure}

In cases when $\beta<\kappa/2$, a second option is to excite the resonator from both ends of the external waveguide, with one frequency aligned to the resonance peak $\mathcal{I}_- = \mathrm{Max}(\mathcal{I})$ and another one aligned such that the two excited fields have equal build-up factors $\tilde{\mathcal{I}}_+=\tilde{\mathcal{I}}_-$ as shown in Fig.~\ref{fig:int} (b). Due to large relative frequency detuning $\delta$ between the two fields, their contributions to the vector light shift, Eq.~(\ref{eqSM:vec}), are of opposite signs and can be summed up incoherently to completely cancel each other. The standing wave pattern in the total scalar shift [Eq.~(\ref{eqSM:vsc})], on the other hand, still remains highly visible. For the example shown in Fig.~\ref{fig:int} (b, d), the two excited fields have unequal intensity buil-up factors, $(\mathcal{I}_+,\mathcal{I}_-) \approx (2.6,1.4)\times 10^3$, visibilities $(V_+,V_-)\approx ( 0.93v, 0.69v)$ and a differential standing-wave phase shift $|\xi_+-\xi_-| \approx 0.49\pi$. The incoherent sum of the scalar light shifts results in a new visibility 
\begin{equation}\label{vprime}
     V' = \frac{\sqrt{(\mathcal{I}_+V_+)^2+(\mathcal{I}_-V_-)^2-2\mathcal{I}_+\mathcal{I}_-V_+V_-\cos|\xi_+-\xi_-|}}{(\mathcal{I}_++\mathcal{I}_-)},
\end{equation}
which gives $V'\approx 0.65v$ and the standing wave pattern remains sufficiently strong.

\subsection{Eliminating standing wave in the scalar potential}
Modifications in the previous schemes allow further elimination of the standing wave potential. We make use of the fact that the standing wave patterns can be created 180 degrees out of phase with respect to each other when we excite the microring from either end of the bus waveguide with exact opposite frequency detuning to the bare resonance $\omega_0$, as shown in Fig.~\ref{fig:int_nolattice} for $\beta > \kappa/2 $ (a,c) and $\beta < \kappa/2 $ (b,d). Since we also have equal energy build-up factors and visibilities, $(\mathcal{I},\tilde{\mathcal{I}},V)$, the standing wave pattern as well as the vector light shift can be fully cancelled, allowing us to create a state-independent, smooth evanescent field potential along the microring that is highly useful in our two-color trapping scheme.

\section{Losses in microring resonators}
\label{appendix:losses}
\subsection{Fundamental limits of the microring platform}
\label{appendix:limit}
Without considering fabrication imperfections, the cooperativity parameter is fundamentally limited by the intrinsic quality factor $Q_\mathrm{i}\sim 1/(Q_\mathrm{a}^{-1} + Q_\mathrm{b}^{-1})$ due to finite material absorption ($Q_\mathrm{a}$) and the bending loss ($Q_\mathrm{b}$). For stoichiometric LPCVD nitride films, it has been estimated that the absorption coefficient $\alpha \ll 1~$dB/m in the near infrared range \cite{Ji:17}. At cesium D1 and D2 wavelengths, for example, we estimate that $Q_\mathrm{a} \gtrsim \ 10^8$ should contribute little to the optical loss in a fabricated microring. On the other hand, we numerically estimate the bending loss from FEM analysis. We have empirically found that $Q_\mathrm{b} \gg 10^8$ when the radius of a microring is beyond 15 micron, as in our case, and the effective refractive index of the resonator mode is $n_\mathrm{eff} > 1.65$, constraining the minimum mode volume to be $V_\mathrm{m} \gtrsim 500\lambda^3$ for an atom trapped around $z_\mathrm{t}\gtrsim 75~$nm ($V_\mathrm{m} \gtrsim 50\lambda^3$ for a solid state emitter at the waveguide surface). Without further considering fabrication imperfections, the fundamental limit for the cooperativity parameter could be as high as $C \sim 10^{-4}Q_\mathrm{a} \gtrsim 10^4$ ($\gtrsim 10^5$ on the waveguide surface).

\subsection{Surface scattering loss}
\label{appendix:loss}
The analysis of surface scattering loss has been greatly discussed in the literature, see \cite{payne1994theoretical,poulton2006radiation,Borselli:05} for example. Here we adopt an analysis similar to \cite{Borselli:05}, but with a number of modifications. We evaluate the surface scattering limited quality factor by calculating 
\begin{equation}
Q_\mathrm{ss}=\frac{\omega U_\mathrm{c}}{P_\mathrm{ss}},
\end{equation}
where $U_\mathrm{c}=\frac{1}{2}\epsilon_0\int \epsilon(\mathbf{r})|\mathbf{E}(\mathbf{r})|^2d^3r$ is the energy stored in the ring, $\epsilon(\mathbf{r})$ and $\mathbf{E}(\mathbf{r})$ are the unperturbed dielectric function and the resonator mode field, respectively, and $P_\mathrm{ss}$ is the radiated power due to surface scattering.

We adopt the volume current method \cite{1074216} to calculate the radiation loss. To leading order, the radiation vector potential is generated by a polarization current density $\mathbf{J}=-i \omega \delta \epsilon \mathbf{E}$ created by the dielectric defects, where $\delta \epsilon(\mathbf{r})$ is the dielectric perturbation function that is non-zero only near the four surfaces of the microring. In the far field, we have
\begin{equation}
\mathbf{A} = \frac{\mu_0}{4\pi} \frac{e^{-i\mathbf{k}\cdot \mathbf{r}}}{r} \int \left[-i \omega \delta \epsilon (\mathbf{r}') \mathbf{E}(\mathbf{r}')\right]e^{-i\mathbf{k}\cdot \mathbf{r}'} d^3r'.\label{vectora}
\end{equation}
The radiation loss can thus be estimated by the time averaged Poynting vector.

We note that the above method works best for a waveguide embedded in a uniform medium \cite{1074216}. In our case, a nitride waveguide on a dioxide substrate embedded in vacuum, an accurate calculation is considerably more complicated due to dielectric discontinuities in the surrounding medium. Here, we neglect multiple reflections and estimate the amount of scattering radiation in the far field (in vacuum) by separately evaluating the contributions from the four surfaces of a microring waveguide. We take $\delta \epsilon_i = \epsilon_0 (\epsilon- \epsilon_i) \Delta_i(\mathbf{r})$, where $\epsilon$ and $\epsilon_i$ are the dielectric constants of nitride and the surrounding dioxide substrate or vacuum, respectively, and $\Delta_i(\mathbf{r})$ represents the distribution function of random irregularities near the $i$-th surface. For the side walls at $\rho_\pm = R \pm W/2$, surface roughness caused by etching imperfections; for top and bottom surfaces at $z=z_{t,b}$, this results from imperfect film growth.

Using, Eqs.~(\ref{vectora}), we could then evaluate 
\begin{align}
P_\mathrm{ss} &= \int\frac{\omega k}{2\mu_0} |\mathbf{r}\times\mathbf{A}|^2 d\Omega \nonumber \\
 &=  \frac{\omega k^3\epsilon_0}{32\pi^2} \sum_{i,\alpha} (\epsilon-\epsilon_i)^2 C_{i,\alpha},\label{poynting_explicit}
\end{align}
where 
\begin{align}
C_{i,\alpha} = & \int d\Omega \int\int \mathcal{E}_\alpha(\mathbf{r'})\mathcal{E}_\alpha^*(\mathbf{r''}) e^{-i\mathbf{k}\cdot(\mathbf{r'}-\mathbf{r''})+im(\phi'-\phi'')}\nonumber \\ 
&\times (\hat{\mathbf{r}}\times \mathbf{\hat{e}}_\alpha')\cdot (\hat{\mathbf{r}}\times \mathbf{\hat{e}}_\alpha'')\Delta_i(\mathbf{r'})\Delta_i(\mathbf{r''})d^3r'd^3r''.\label{ci}
\end{align}
The random roughness in the second line of the integral varies at very small length scale $\ll \lambda$, as suggested by our AFM measurements. Thus, the electric field related terms in the first and the second lines above can be considered slow-varying. The integration over large ring surfaces should sample many local patches of irregularities, each weighed by similar electric field value and polarization orientation. We may thus replace $\Delta_i(\mathbf{r'})\Delta_i(\mathbf{r''})$ with an ensemble averaged two-point correlation function $\langle \Delta_i(\mathbf{r'})\Delta_i(\mathbf{r''})\rangle$, which can be determined from the AFM measurements. We approximate the two-point correlation with a Gaussian form  
\begin{align}
    \langle\Delta&_i(\mathbf{r}')\Delta_{i}(\mathbf{r}'')\rangle \approx \nonumber \\
     &\sigma_i^2e^{-\frac{(\mathbf{r'}-\mathbf{r''})^2}{L_i^2}}\delta(z'-z_i)\delta(z''-z_i)\Theta(\rho')\Theta(\rho'') \label{gaussiantb}
\end{align}
for top and bottom surfaces ($i=t,b$) and, similarly,
\begin{align}
    \langle\Delta&_i(\mathbf{r}')\Delta_{i}(\mathbf{r}'')\rangle \approx   \nonumber \\
     &\sigma_i^2e^{-\frac{R_i^2(\phi'-\phi'')^2}{L_i^2}}\delta(\rho'-\rho_i)\delta(\rho''-\rho_i) \Theta(z')\Theta(z'')\label{gaussiansw}
\end{align}
for the side walls ($i=\pm$). In the above, $\sigma_i$ and $L_i$ are the root-mean-squared roughness and the correlation length, respectively. $\delta(x)$ is the Dirac delta function, and $\Theta(x)=1$ for $x$ lying within the range of the (perfect) ring waveguide and $\Theta(x)=0$ otherwise.

Plugging Eq.~(\ref{gaussiantb}) into Eq.~(\ref{ci}) to evaluate loss contribution from top and bottom roughness, we obtain
\begin{equation}
     C_{i,\alpha} \approx \sigma_i^2 \int\int d\rho'd\rho''\rho'^2 |\mathcal{E}_\alpha(\rho',z_i)|^2 e^{-\frac{(\rho'-\rho'')^2}{L_i^2}} \Phi_\alpha(\rho'),\label{ccl}
\end{equation}
where, due to the short correlation length $L_i\ll \lambda$, we can simplify the azymuthal part of the integral by taking $\rho'' \approx \rho'$ and arrive at the following
\begin{align}
     \Phi_\alpha = & \int d\Omega \int d\phi' d\phi'' (\hat{\mathbf{r}}\times \mathbf{\hat{e}}_\alpha')\cdot (\hat{\mathbf{r}}\times \mathbf{\hat{e}}_\alpha'') \times \nonumber \\ & e^{im(\phi'-\phi'')}e^{-ik\frac{\rho}{r}\rho'(\cos\phi'-\cos\phi'')}e^{-\frac{\rho'^2(\phi'-\phi'')^2}{L_i^2}} \nonumber \\
    \sim & 4\pi^{5/2} \eta \frac{L_i}{\rho'}.\label{approx}
\end{align}
In the above, we used the fact that the integrant is none-vanishing only when $|\phi'-\phi''| \lesssim L_i/R \ll 1$ and $m|\phi'-\phi''| \sim kL_i \lesssim 1$. Here $\eta = \frac{4}{3}$ is a geometric radiation parameter due to mode-field polarizations coupled to that of freespace radiation modes \cite{Borselli:05}; We note that $\eta$ is polarization independent, different from the result of \cite{Borselli:05}, because the surface scatterers are approximately spherically symmetric ($\sigma_i, L_i \ll \lambda$) in the sense of radiation at farfield.

Plugging Eq.~(\ref{approx}) into Eq.~(\ref{ccl}), we arrive at 
\begin{align}
     C_{\mathrm{t(b)},\alpha} \approx & 4\pi^{5/2}\sigma_{\mathrm{t(b)}}^2 L_{\mathrm{t(b)}}^2 R W \eta |\bar{\mathcal{E}}_{\mathrm{t(b)},\alpha}|^2 \nonumber \\
     = & \frac{16}{3}\pi^{5/2}V_{\mathrm{t(b)}}^2 |\bar{\mathcal{E}}_{\mathrm{t(b)},\alpha}|^2,\label{ctb}
\end{align}
where $|\bar{\mathcal{E}}_{\mathrm{t(b)},\alpha}|^2=\frac{1}{RW}\int_{\rho_-}^{\rho_+}\rho'|\mathcal{E}_\alpha(\rho',z_{\mathrm{t(b)}})|^2d\rho'$ is the averaged mode field and $V_{\mathrm{t(b)}}\equiv \sigma_{\mathrm{t(b)}}L_{\mathrm{t(b)}}\sqrt{RW}$ is the effective volume of the scatterers.

For scattering contributions due to side wall roughness, we adopt similar procedures and obtain
\begin{equation}
     C_{\pm,\alpha} \approx \frac{16}{3}\pi^{5/2}V_\pm^2 |\bar{\mathcal{E}}_{\pm,\alpha}|^2,
     \label{csw}
\end{equation}
where the effective volume is $V_\pm = \sigma_\pm H \sqrt{L_\pm \rho_\pm}$, the effective mode field $|\bar{\mathcal{E}}_{\pm,\alpha}|^2 = \frac{1}{\eta H^2}\int_{-1}^{1}   |\tilde{\mathcal{E}}_{\pm,\alpha}(\tilde{k})|^2 \eta_\alpha(\tilde{k}) d\tilde{k}$ and $\tilde{\mathcal{E}}_{\pm,\alpha}(\tilde{k}) = \int_0^H \mathcal{E}_\alpha(\rho_\pm,z')e^{-ik\tilde{k}z' }dz'$. Here, $\eta_\alpha$ is polarization dependent due to the geometric shape of the side wall roughness, with $ \eta_{\rho(\phi)}(\tilde{k}) = \frac{1}{2}(1+\tilde{k}^2)$ and $\eta_z = 1-\tilde{k}^2$.

We note that the apparent difference between the mode field contributions in Eqs.~(\ref{ctb}) and (\ref{csw}) is due to the effective volume of the surface and side wall scatterers. At the side walls, because the vertical length of the edge roughness is about the thickness $H$ of the waveguide and is rather comparable to the wavelength, interference effect manifests and modifies the scattering contribution from the mode field $|\bar{\mathcal{E}}_{\pm,\alpha}|^2$. If $kH \ll 1$, $|\bar{\mathcal{E}}_{\pm,\alpha}|^2 \approx |\int_0^H \mathcal{E}_\alpha(R_\pm,z') dz'/H|^2$ gives the averaged mode field squared at the side walls. 

We then evaluate the scattering-loss quality factor as
\begin{align}
     Q_\mathrm{ss} = & \frac{32\pi^2 U_\mathrm{c}}{k^3\epsilon_0 \sum_{i,\alpha} \Delta \epsilon_i^2C_{i,\alpha}}\\
      = & \frac{3\lambda^3}{8\pi^{7/2}\sum_{i,\alpha} \Delta \epsilon_i^2 V_i^2 |\bar{u}_{i,\alpha}|^2}, \label{qss}
\end{align}
where $\Delta \epsilon_i = \epsilon - \epsilon_i$ and define
\begin{align}
|\bar{u}_{i,\alpha}|^2 \equiv & \frac{|\bar{\mathcal{E}}_{i,\alpha}|^2}{\int \epsilon(\mathbf{r})|\mathbf{E}(\mathbf{r})|^2d^3r}\\
\approx & \frac{|\bar{\mathcal{E}}_{i,\alpha}|^2}{2\pi R \int \epsilon(\mathbf{r})\sum_\alpha|\mathcal{E}_\alpha(\rho, z)|^2d\rho dz}
\end{align}
as the normalized, weighted mode field energy density.

In the main text we optimize $Q/V_\mathrm{m}$ by calculating $Q=1/(Q_\mathrm{b}^{-1} + Q_\mathrm{ss}^{-1})$. From Eq.~(\ref{qss}) with given roughness parameters, it is clear that $Q_\mathrm{ss}$ can be made higher by increasing the degree of mode confinement, which requires increasing the cross-section of the microring ($H,W$) to mitigate the surface scattering loss. However, this will be constrained by the desire to decrease the mode volume, that is, to increase the mode field strength at the atomic trap location $|\mathcal{E}(\rho_\mathrm{t},z_\mathrm{t})|^2$. Moreover, the radius of the microring cannot be reduced indefinitely because of the increased bending loss and the surface scattering loss at the sidewall (the guided mode shifts toward the outer edge of the microring, as shown in Fig.~\ref{fig:TETMEi}). An optimized geometry balances the requirement for proper mode confinement and small mode volume to achieve high $Q/V_\mathrm{m}$.

\newpage

%% file: Microring.bbl
\begin{thebibliography}{54}
\expandafter\ifx\csname natexlab\endcsname\relax\def\natexlab#1{#1}\fi
\expandafter\ifx\csname bibnamefont\endcsname\relax
  \def\bibnamefont#1{#1}\fi
\expandafter\ifx\csname bibfnamefont\endcsname\relax
  \def\bibfnamefont#1{#1}\fi
\expandafter\ifx\csname citenamefont\endcsname\relax
  \def\citenamefont#1{#1}\fi
\expandafter\ifx\csname url\endcsname\relax
  \def\url#1{\texttt{#1}}\fi
\expandafter\ifx\csname urlprefix\endcsname\relax\def\urlprefix{URL }\fi
\providecommand{\bibinfo}[2]{#2}
\providecommand{\eprint}[2][]{\url{#2}}

\bibitem[{\citenamefont{O'brien et~al.}(2009)\citenamefont{O'brien, Furusawa,
  and Vu{\v{c}}kovi{\'c}}}]{o2009photonic}
\bibinfo{author}{\bibfnamefont{J.~L.} \bibnamefont{O'brien}},
  \bibinfo{author}{\bibfnamefont{A.}~\bibnamefont{Furusawa}}, \bibnamefont{and}
  \bibinfo{author}{\bibfnamefont{J.}~\bibnamefont{Vu{\v{c}}kovi{\'c}}},
  \bibinfo{journal}{Nature Photonics} \textbf{\bibinfo{volume}{3}},
  \bibinfo{pages}{687} (\bibinfo{year}{2009}).

\bibitem[{\citenamefont{Cirac and Kimble}(2017)}]{Cirac_photonics_2017}
\bibinfo{author}{\bibfnamefont{J.~I.} \bibnamefont{Cirac}} \bibnamefont{and}
  \bibinfo{author}{\bibfnamefont{H.~J.} \bibnamefont{Kimble}},
  \bibinfo{journal}{Nature Photonics} \textbf{\bibinfo{volume}{11}},
  \bibinfo{pages}{18 EP } (\bibinfo{year}{2017}).

\bibitem[{\citenamefont{Chang et~al.}(2018)\citenamefont{Chang, Douglas,
  Gonz\'alez-Tudela, Hung, and Kimble}}]{ChangRMP2018}
\bibinfo{author}{\bibfnamefont{D.~E.} \bibnamefont{Chang}},
  \bibinfo{author}{\bibfnamefont{J.~S.} \bibnamefont{Douglas}},
  \bibinfo{author}{\bibfnamefont{A.}~\bibnamefont{Gonz\'alez-Tudela}},
  \bibinfo{author}{\bibfnamefont{C.-L.} \bibnamefont{Hung}}, \bibnamefont{and}
  \bibinfo{author}{\bibfnamefont{H.~J.} \bibnamefont{Kimble}},
  \bibinfo{journal}{Rev. Mod. Phys.} \textbf{\bibinfo{volume}{90}},
  \bibinfo{pages}{031002} (\bibinfo{year}{2018}).

\bibitem[{\citenamefont{Vetsch et~al.}(2010)\citenamefont{Vetsch, Reitz, Sague,
  Schmidt, Dawkins, and Rauschenbeutel}}]{vetsch_optical_2010}
\bibinfo{author}{\bibfnamefont{E.}~\bibnamefont{Vetsch}},
  \bibinfo{author}{\bibfnamefont{D.}~\bibnamefont{Reitz}},
  \bibinfo{author}{\bibfnamefont{G.}~\bibnamefont{Sague}},
  \bibinfo{author}{\bibfnamefont{R.}~\bibnamefont{Schmidt}},
  \bibinfo{author}{\bibfnamefont{S.~T.} \bibnamefont{Dawkins}},
  \bibnamefont{and}
  \bibinfo{author}{\bibfnamefont{A.}~\bibnamefont{Rauschenbeutel}},
  \bibinfo{journal}{Phys. Rev. Lett.} \textbf{\bibinfo{volume}{104}},
  \bibinfo{pages}{203603} (\bibinfo{year}{2010}).

\bibitem[{\citenamefont{Goban et~al.}(2012)\citenamefont{Goban, Choi, Alton,
  Ding, Lacroute, Pototschnig, Thiele, Stern, and
  Kimble}}]{goban_demonstration_2012}
\bibinfo{author}{\bibfnamefont{A.}~\bibnamefont{Goban}},
  \bibinfo{author}{\bibfnamefont{K.~S.} \bibnamefont{Choi}},
  \bibinfo{author}{\bibfnamefont{D.~J.} \bibnamefont{Alton}},
  \bibinfo{author}{\bibfnamefont{D.}~\bibnamefont{Ding}},
  \bibinfo{author}{\bibfnamefont{C.}~\bibnamefont{Lacroute}},
  \bibinfo{author}{\bibfnamefont{M.}~\bibnamefont{Pototschnig}},
  \bibinfo{author}{\bibfnamefont{T.}~\bibnamefont{Thiele}},
  \bibinfo{author}{\bibfnamefont{N.~P.} \bibnamefont{Stern}}, \bibnamefont{and}
  \bibinfo{author}{\bibfnamefont{H.~J.} \bibnamefont{Kimble}},
  \bibinfo{journal}{Phys. Rev. Lett.} \textbf{\bibinfo{volume}{109}},
  \bibinfo{pages}{033603} (\bibinfo{year}{2012}).

\bibitem[{\citenamefont{Kato and Aoki}(2015)}]{kato_strong_2015}
\bibinfo{author}{\bibfnamefont{S.}~\bibnamefont{Kato}} \bibnamefont{and}
  \bibinfo{author}{\bibfnamefont{T.}~\bibnamefont{Aoki}},
  \bibinfo{journal}{Physical Review Letters} \textbf{\bibinfo{volume}{115}},
  \bibinfo{pages}{093603} (\bibinfo{year}{2015}).

\bibitem[{\citenamefont{Sorensen et~al.}(2016)\citenamefont{Sorensen, Beguin,
  Kluge, Iakoupov, Sorensen, Muller, Polzik, and
  Appel}}]{sorensen_coherent_2016}
\bibinfo{author}{\bibfnamefont{H.}~\bibnamefont{Sorensen}},
  \bibinfo{author}{\bibfnamefont{J.-B.} \bibnamefont{Beguin}},
  \bibinfo{author}{\bibfnamefont{K.}~\bibnamefont{Kluge}},
  \bibinfo{author}{\bibfnamefont{I.}~\bibnamefont{Iakoupov}},
  \bibinfo{author}{\bibfnamefont{A.}~\bibnamefont{Sorensen}},
  \bibinfo{author}{\bibfnamefont{J.}~\bibnamefont{Muller}},
  \bibinfo{author}{\bibfnamefont{E.}~\bibnamefont{Polzik}}, \bibnamefont{and}
  \bibinfo{author}{\bibfnamefont{J.}~\bibnamefont{Appel}},
  \bibinfo{journal}{Physical Review Letters} \textbf{\bibinfo{volume}{117}},
  \bibinfo{pages}{133604} (\bibinfo{year}{2016}).

\bibitem[{\citenamefont{Corzo et~al.}(2016)\citenamefont{Corzo, Gouraud,
  Chandra, Goban, Sheremet, Kupriyanov, and Laurat}}]{corzo_large_2016}
\bibinfo{author}{\bibfnamefont{N.~V.} \bibnamefont{Corzo}},
  \bibinfo{author}{\bibfnamefont{B.}~\bibnamefont{Gouraud}},
  \bibinfo{author}{\bibfnamefont{A.}~\bibnamefont{Chandra}},
  \bibinfo{author}{\bibfnamefont{A.}~\bibnamefont{Goban}},
  \bibinfo{author}{\bibfnamefont{A.~S.} \bibnamefont{Sheremet}},
  \bibinfo{author}{\bibfnamefont{D.~V.} \bibnamefont{Kupriyanov}},
  \bibnamefont{and} \bibinfo{author}{\bibfnamefont{J.}~\bibnamefont{Laurat}},
  \bibinfo{journal}{Physical Review Letters} \textbf{\bibinfo{volume}{117}},
  \bibinfo{pages}{133603} (\bibinfo{year}{2016}).

\bibitem[{\citenamefont{Goban et~al.}(2015)\citenamefont{Goban, Hung, Hood, Yu,
  Muniz, Painter, and Kimble}}]{goban_superradiance_2015}
\bibinfo{author}{\bibfnamefont{A.}~\bibnamefont{Goban}},
  \bibinfo{author}{\bibfnamefont{C.-L.} \bibnamefont{Hung}},
  \bibinfo{author}{\bibfnamefont{J.}~\bibnamefont{Hood}},
  \bibinfo{author}{\bibfnamefont{S.-P.} \bibnamefont{Yu}},
  \bibinfo{author}{\bibfnamefont{J.}~\bibnamefont{Muniz}},
  \bibinfo{author}{\bibfnamefont{O.}~\bibnamefont{Painter}}, \bibnamefont{and}
  \bibinfo{author}{\bibfnamefont{H.}~\bibnamefont{Kimble}},
  \bibinfo{journal}{Physical Review Letters} \textbf{\bibinfo{volume}{115}},
  \bibinfo{pages}{063601} (\bibinfo{year}{2015}).

\bibitem[{\citenamefont{Thompson et~al.}(2013)\citenamefont{Thompson, Tiecke,
  de~Leon, Feist, Akimov, Gullans, Zibrov, Vuletic, and
  Lukin}}]{thompson_coupling_2013}
\bibinfo{author}{\bibfnamefont{J.~D.} \bibnamefont{Thompson}},
  \bibinfo{author}{\bibfnamefont{T.~G.} \bibnamefont{Tiecke}},
  \bibinfo{author}{\bibfnamefont{N.~P.} \bibnamefont{de~Leon}},
  \bibinfo{author}{\bibfnamefont{J.}~\bibnamefont{Feist}},
  \bibinfo{author}{\bibfnamefont{A.~V.} \bibnamefont{Akimov}},
  \bibinfo{author}{\bibfnamefont{M.}~\bibnamefont{Gullans}},
  \bibinfo{author}{\bibfnamefont{A.~S.} \bibnamefont{Zibrov}},
  \bibinfo{author}{\bibfnamefont{V.}~\bibnamefont{Vuletic}}, \bibnamefont{and}
  \bibinfo{author}{\bibfnamefont{M.~D.} \bibnamefont{Lukin}},
  \bibinfo{journal}{Science} \textbf{\bibinfo{volume}{340}},
  \bibinfo{pages}{1202} (\bibinfo{year}{2013}), ISSN \bibinfo{issn}{0036-8075,
  1095-9203}.

\bibitem[{\citenamefont{Tiecke et~al.}(2014)\citenamefont{Tiecke, Thompson,
  de~Leon, Liu, Vuletic, and Lukin}}]{tiecke_nanophotonic_2014}
\bibinfo{author}{\bibfnamefont{T.~G.} \bibnamefont{Tiecke}},
  \bibinfo{author}{\bibfnamefont{J.~D.} \bibnamefont{Thompson}},
  \bibinfo{author}{\bibfnamefont{N.~P.} \bibnamefont{de~Leon}},
  \bibinfo{author}{\bibfnamefont{L.~R.} \bibnamefont{Liu}},
  \bibinfo{author}{\bibfnamefont{V.}~\bibnamefont{Vuletic}}, \bibnamefont{and}
  \bibinfo{author}{\bibfnamefont{M.~D.} \bibnamefont{Lukin}},
  \bibinfo{journal}{Nature} \textbf{\bibinfo{volume}{508}},
  \bibinfo{pages}{241} (\bibinfo{year}{2014}), ISSN \bibinfo{issn}{0028-0836}.

\bibitem[{\citenamefont{Grimm et~al.}(2000)\citenamefont{Grimm,
  Weidem{\"u}ller, and Ovchinnikov}}]{grimm2000optical}
\bibinfo{author}{\bibfnamefont{R.}~\bibnamefont{Grimm}},
  \bibinfo{author}{\bibfnamefont{M.}~\bibnamefont{Weidem{\"u}ller}},
  \bibnamefont{and} \bibinfo{author}{\bibfnamefont{Y.~B.}
  \bibnamefont{Ovchinnikov}}, in \emph{\bibinfo{booktitle}{Advances in atomic,
  molecular, and optical physics}} (\bibinfo{publisher}{Elsevier},
  \bibinfo{year}{2000}), vol.~\bibinfo{volume}{42}, pp.
  \bibinfo{pages}{95--170}.

\bibitem[{\citenamefont{Le~Kien et~al.}(2004)\citenamefont{Le~Kien, Balykin,
  and Hakuta}}]{le_kien_atom_2004}
\bibinfo{author}{\bibfnamefont{F.}~\bibnamefont{Le~Kien}},
  \bibinfo{author}{\bibfnamefont{V.~I.} \bibnamefont{Balykin}},
  \bibnamefont{and} \bibinfo{author}{\bibfnamefont{K.}~\bibnamefont{Hakuta}},
  \bibinfo{journal}{Physical Review A} \textbf{\bibinfo{volume}{70}},
  \bibinfo{pages}{063403} (\bibinfo{year}{2004}).

\bibitem[{\citenamefont{Balykin et~al.}(2004)\citenamefont{Balykin, Hakuta,
  Le~Kien, Liang, and Morinaga}}]{balykin2004atom}
\bibinfo{author}{\bibfnamefont{V.}~\bibnamefont{Balykin}},
  \bibinfo{author}{\bibfnamefont{K.}~\bibnamefont{Hakuta}},
  \bibinfo{author}{\bibfnamefont{F.}~\bibnamefont{Le~Kien}},
  \bibinfo{author}{\bibfnamefont{J.}~\bibnamefont{Liang}}, \bibnamefont{and}
  \bibinfo{author}{\bibfnamefont{M.}~\bibnamefont{Morinaga}},
  \bibinfo{journal}{Physical Review A} \textbf{\bibinfo{volume}{70}},
  \bibinfo{pages}{011401} (\bibinfo{year}{2004}).

\bibitem[{\citenamefont{Lacro{\^u}te et~al.}(2012)\citenamefont{Lacro{\^u}te,
  Choi, Goban, Alton, Ding, Stern, and Kimble}}]{lacroute2012state}
\bibinfo{author}{\bibfnamefont{C.}~\bibnamefont{Lacro{\^u}te}},
  \bibinfo{author}{\bibfnamefont{K.}~\bibnamefont{Choi}},
  \bibinfo{author}{\bibfnamefont{A.}~\bibnamefont{Goban}},
  \bibinfo{author}{\bibfnamefont{D.}~\bibnamefont{Alton}},
  \bibinfo{author}{\bibfnamefont{D.}~\bibnamefont{Ding}},
  \bibinfo{author}{\bibfnamefont{N.}~\bibnamefont{Stern}}, \bibnamefont{and}
  \bibinfo{author}{\bibfnamefont{H.}~\bibnamefont{Kimble}},
  \bibinfo{journal}{New Journal of Physics} \textbf{\bibinfo{volume}{14}},
  \bibinfo{pages}{023056} (\bibinfo{year}{2012}).

\bibitem[{\citenamefont{P{\'e}rez-R{\'\i}os
  et~al.}(2017)\citenamefont{P{\'e}rez-R{\'\i}os, Kim, and
  Hung}}]{Rios_molecule_2017}
\bibinfo{author}{\bibfnamefont{J.}~\bibnamefont{P{\'e}rez-R{\'\i}os}},
  \bibinfo{author}{\bibfnamefont{M.~E.} \bibnamefont{Kim}}, \bibnamefont{and}
  \bibinfo{author}{\bibfnamefont{C.-L.} \bibnamefont{Hung}},
  \bibinfo{journal}{New Journal of Physics} \textbf{\bibinfo{volume}{19}},
  \bibinfo{pages}{123035} (\bibinfo{year}{2017}).

\bibitem[{\citenamefont{Hung et~al.}(2013)\citenamefont{Hung, Meenehan, Chang,
  Painter, and Kimble}}]{hung_trapped_2013}
\bibinfo{author}{\bibfnamefont{C.-L.} \bibnamefont{Hung}},
  \bibinfo{author}{\bibfnamefont{S.~M.} \bibnamefont{Meenehan}},
  \bibinfo{author}{\bibfnamefont{D.~E.} \bibnamefont{Chang}},
  \bibinfo{author}{\bibfnamefont{O.}~\bibnamefont{Painter}}, \bibnamefont{and}
  \bibinfo{author}{\bibfnamefont{H.~J.} \bibnamefont{Kimble}},
  \bibinfo{journal}{New Journal of Physics} \textbf{\bibinfo{volume}{15}},
  \bibinfo{pages}{083026} (\bibinfo{year}{2013}), ISSN
  \bibinfo{issn}{1367-2630}.

\bibitem[{\citenamefont{Gonzalez-Tudela
  et~al.}(2015)\citenamefont{Gonzalez-Tudela, Hung, Chang, Cirac, and
  Kimble}}]{gonzalez-tudela_subwavelength_2015}
\bibinfo{author}{\bibfnamefont{A.}~\bibnamefont{Gonzalez-Tudela}},
  \bibinfo{author}{\bibfnamefont{C.-L.} \bibnamefont{Hung}},
  \bibinfo{author}{\bibfnamefont{D.~E.} \bibnamefont{Chang}},
  \bibinfo{author}{\bibfnamefont{J.~I.} \bibnamefont{Cirac}}, \bibnamefont{and}
  \bibinfo{author}{\bibfnamefont{H.~J.} \bibnamefont{Kimble}},
  \bibinfo{journal}{Nature Photonics} \textbf{\bibinfo{volume}{9}},
  \bibinfo{pages}{320} (\bibinfo{year}{2015}), ISSN \bibinfo{issn}{1749-4885}.

\bibitem[{\citenamefont{Douglas et~al.}(2015)\citenamefont{Douglas, Habibian,
  Hung, Gorshkov, Kimble, and Chang}}]{douglas_quantum_2015}
\bibinfo{author}{\bibfnamefont{J.~S.} \bibnamefont{Douglas}},
  \bibinfo{author}{\bibfnamefont{H.}~\bibnamefont{Habibian}},
  \bibinfo{author}{\bibfnamefont{C.-L.} \bibnamefont{Hung}},
  \bibinfo{author}{\bibfnamefont{A.}~\bibnamefont{Gorshkov}},
  \bibinfo{author}{\bibfnamefont{H.~J.} \bibnamefont{Kimble}},
  \bibnamefont{and} \bibinfo{author}{\bibfnamefont{D.~E.} \bibnamefont{Chang}},
  \bibinfo{journal}{Nature Photonics} \textbf{\bibinfo{volume}{9}},
  \bibinfo{pages}{326} (\bibinfo{year}{2015}).

\bibitem[{\citenamefont{Hood et~al.}(2016)\citenamefont{Hood, Goban,
  Asenjo-Garcia, Lu, Yu, Chang, and Kimble}}]{hood2016atom}
\bibinfo{author}{\bibfnamefont{J.~D.} \bibnamefont{Hood}},
  \bibinfo{author}{\bibfnamefont{A.}~\bibnamefont{Goban}},
  \bibinfo{author}{\bibfnamefont{A.}~\bibnamefont{Asenjo-Garcia}},
  \bibinfo{author}{\bibfnamefont{M.}~\bibnamefont{Lu}},
  \bibinfo{author}{\bibfnamefont{S.-P.} \bibnamefont{Yu}},
  \bibinfo{author}{\bibfnamefont{D.~E.} \bibnamefont{Chang}}, \bibnamefont{and}
  \bibinfo{author}{\bibfnamefont{H.}~\bibnamefont{Kimble}},
  \bibinfo{journal}{Proceedings of the National Academy of Sciences}
  \textbf{\bibinfo{volume}{113}}, \bibinfo{pages}{10507}
  (\bibinfo{year}{2016}).

\bibitem[{\citenamefont{Aoki et~al.}(2006)\citenamefont{Aoki, Dayan, Wilcut,
  Bowen, Parkins, Kippenberg, Vahala, and Kimble}}]{aoki2006observation}
\bibinfo{author}{\bibfnamefont{T.}~\bibnamefont{Aoki}},
  \bibinfo{author}{\bibfnamefont{B.}~\bibnamefont{Dayan}},
  \bibinfo{author}{\bibfnamefont{E.}~\bibnamefont{Wilcut}},
  \bibinfo{author}{\bibfnamefont{W.~P.} \bibnamefont{Bowen}},
  \bibinfo{author}{\bibfnamefont{A.~S.} \bibnamefont{Parkins}},
  \bibinfo{author}{\bibfnamefont{T.}~\bibnamefont{Kippenberg}},
  \bibinfo{author}{\bibfnamefont{K.}~\bibnamefont{Vahala}}, \bibnamefont{and}
  \bibinfo{author}{\bibfnamefont{H.}~\bibnamefont{Kimble}},
  \bibinfo{journal}{Nature} \textbf{\bibinfo{volume}{443}},
  \bibinfo{pages}{671} (\bibinfo{year}{2006}).

\bibitem[{\citenamefont{O’Shea et~al.}(2013)\citenamefont{O’Shea, Junge,
  Volz, and Rauschenbeutel}}]{o2013fiber}
\bibinfo{author}{\bibfnamefont{D.}~\bibnamefont{O’Shea}},
  \bibinfo{author}{\bibfnamefont{C.}~\bibnamefont{Junge}},
  \bibinfo{author}{\bibfnamefont{J.}~\bibnamefont{Volz}}, \bibnamefont{and}
  \bibinfo{author}{\bibfnamefont{A.}~\bibnamefont{Rauschenbeutel}},
  \bibinfo{journal}{Physical review letters} \textbf{\bibinfo{volume}{111}},
  \bibinfo{pages}{193601} (\bibinfo{year}{2013}).

\bibitem[{\citenamefont{Shomroni et~al.}(2014)\citenamefont{Shomroni,
  Rosenblum, Lovsky, Bechler, Guendelman, and Dayan}}]{shomroni2014all}
\bibinfo{author}{\bibfnamefont{I.}~\bibnamefont{Shomroni}},
  \bibinfo{author}{\bibfnamefont{S.}~\bibnamefont{Rosenblum}},
  \bibinfo{author}{\bibfnamefont{Y.}~\bibnamefont{Lovsky}},
  \bibinfo{author}{\bibfnamefont{O.}~\bibnamefont{Bechler}},
  \bibinfo{author}{\bibfnamefont{G.}~\bibnamefont{Guendelman}},
  \bibnamefont{and} \bibinfo{author}{\bibfnamefont{B.}~\bibnamefont{Dayan}},
  \bibinfo{journal}{Science} \textbf{\bibinfo{volume}{345}},
  \bibinfo{pages}{903} (\bibinfo{year}{2014}).

\bibitem[{\citenamefont{Barclay et~al.}(2006)\citenamefont{Barclay, Srinivasan,
  Painter, Lev, and Mabuchi}}]{barclay2006integration}
\bibinfo{author}{\bibfnamefont{P.~E.} \bibnamefont{Barclay}},
  \bibinfo{author}{\bibfnamefont{K.}~\bibnamefont{Srinivasan}},
  \bibinfo{author}{\bibfnamefont{O.}~\bibnamefont{Painter}},
  \bibinfo{author}{\bibfnamefont{B.}~\bibnamefont{Lev}}, \bibnamefont{and}
  \bibinfo{author}{\bibfnamefont{H.}~\bibnamefont{Mabuchi}},
  \bibinfo{journal}{Applied physics letters} \textbf{\bibinfo{volume}{89}},
  \bibinfo{pages}{131108} (\bibinfo{year}{2006}).

\bibitem[{\citenamefont{Alton}(2013)}]{alton2013interacting}
\bibinfo{author}{\bibfnamefont{D.~J.} \bibnamefont{Alton}}, Ph.D. thesis,
  \bibinfo{school}{California Institute of Technology} (\bibinfo{year}{2013}).

\bibitem[{\citenamefont{Xuan et~al.}(2016)\citenamefont{Xuan, Liu, Varghese,
  Metcalf, Xue, Wang, Han, Jaramillo-Villegas, Noman, Wang et~al.}}]{Xuan:16}
\bibinfo{author}{\bibfnamefont{Y.}~\bibnamefont{Xuan}},
  \bibinfo{author}{\bibfnamefont{Y.}~\bibnamefont{Liu}},
  \bibinfo{author}{\bibfnamefont{L.~T.} \bibnamefont{Varghese}},
  \bibinfo{author}{\bibfnamefont{A.~J.} \bibnamefont{Metcalf}},
  \bibinfo{author}{\bibfnamefont{X.}~\bibnamefont{Xue}},
  \bibinfo{author}{\bibfnamefont{P.-H.} \bibnamefont{Wang}},
  \bibinfo{author}{\bibfnamefont{K.}~\bibnamefont{Han}},
  \bibinfo{author}{\bibfnamefont{J.~A.} \bibnamefont{Jaramillo-Villegas}},
  \bibinfo{author}{\bibfnamefont{A.~A.} \bibnamefont{Noman}},
  \bibinfo{author}{\bibfnamefont{C.}~\bibnamefont{Wang}}, \bibnamefont{et~al.},
  \bibinfo{journal}{Optica} \textbf{\bibinfo{volume}{3}}, \bibinfo{pages}{1171}
  (\bibinfo{year}{2016}).

\bibitem[{\citenamefont{Ji et~al.}(2017)\citenamefont{Ji, Barbosa, Roberts,
  Dutt, Cardenas, Okawachi, Bryant, Gaeta, and Lipson}}]{Ji:17}
\bibinfo{author}{\bibfnamefont{X.}~\bibnamefont{Ji}},
  \bibinfo{author}{\bibfnamefont{F.~A.~S.} \bibnamefont{Barbosa}},
  \bibinfo{author}{\bibfnamefont{S.~P.} \bibnamefont{Roberts}},
  \bibinfo{author}{\bibfnamefont{A.}~\bibnamefont{Dutt}},
  \bibinfo{author}{\bibfnamefont{J.}~\bibnamefont{Cardenas}},
  \bibinfo{author}{\bibfnamefont{Y.}~\bibnamefont{Okawachi}},
  \bibinfo{author}{\bibfnamefont{A.}~\bibnamefont{Bryant}},
  \bibinfo{author}{\bibfnamefont{A.~L.} \bibnamefont{Gaeta}}, \bibnamefont{and}
  \bibinfo{author}{\bibfnamefont{M.}~\bibnamefont{Lipson}},
  \bibinfo{journal}{Optica} \textbf{\bibinfo{volume}{4}}, \bibinfo{pages}{619}
  (\bibinfo{year}{2017}).

\bibitem[{\citenamefont{Kaufmann et~al.}(2018)\citenamefont{Kaufmann, Ji, Luke,
  Lipson, and Ramelow}}]{Kaufmann:18}
\bibinfo{author}{\bibfnamefont{P.}~\bibnamefont{Kaufmann}},
  \bibinfo{author}{\bibfnamefont{X.}~\bibnamefont{Ji}},
  \bibinfo{author}{\bibfnamefont{K.}~\bibnamefont{Luke}},
  \bibinfo{author}{\bibfnamefont{M.}~\bibnamefont{Lipson}}, \bibnamefont{and}
  \bibinfo{author}{\bibfnamefont{S.}~\bibnamefont{Ramelow}}, in
  \emph{\bibinfo{booktitle}{Conference on Lasers and Electro-Optics}}
  (\bibinfo{publisher}{Optical Society of America}, \bibinfo{year}{2018}), p.
  \bibinfo{pages}{JTu2A.72}.

\bibitem[{\citenamefont{Kim et~al.}(2019)\citenamefont{Kim, Chang, Fields,
  Chen, and Hung}}]{kim2019}
\bibinfo{author}{\bibfnamefont{M.~E.} \bibnamefont{Kim}},
  \bibinfo{author}{\bibfnamefont{T.-H.} \bibnamefont{Chang}},
  \bibinfo{author}{\bibfnamefont{B.~M.} \bibnamefont{Fields}},
  \bibinfo{author}{\bibfnamefont{C.-A.} \bibnamefont{Chen}}, \bibnamefont{and}
  \bibinfo{author}{\bibfnamefont{C.-L.} \bibnamefont{Hung}},
  \bibinfo{journal}{Nature Communications} \textbf{\bibinfo{volume}{10}},
  \bibinfo{pages}{1647} (\bibinfo{year}{2019}).

\bibitem[{cit({\natexlab{a}})}]{cite:lumerical}
\bibinfo{note}{\url{https://www.lumerical.com}}.

\bibitem[{\citenamefont{Srinivasan and Painter}(2007)}]{Srinivasan2007}
\bibinfo{author}{\bibfnamefont{K.}~\bibnamefont{Srinivasan}} \bibnamefont{and}
  \bibinfo{author}{\bibfnamefont{O.}~\bibnamefont{Painter}},
  \bibinfo{journal}{Phys. Rev. A} \textbf{\bibinfo{volume}{75}},
  \bibinfo{pages}{023814} (\bibinfo{year}{2007}).

\bibitem[{\citenamefont{Roberts et~al.}(2017)\citenamefont{Roberts, Ji,
  Cardenas, Bryant, and Lipson}}]{Roberts:17}
\bibinfo{author}{\bibfnamefont{S.~P.} \bibnamefont{Roberts}},
  \bibinfo{author}{\bibfnamefont{X.}~\bibnamefont{Ji}},
  \bibinfo{author}{\bibfnamefont{J.}~\bibnamefont{Cardenas}},
  \bibinfo{author}{\bibfnamefont{A.}~\bibnamefont{Bryant}}, \bibnamefont{and}
  \bibinfo{author}{\bibfnamefont{M.}~\bibnamefont{Lipson}}, in
  \emph{\bibinfo{booktitle}{Conference on Lasers and Electro-Optics}}
  (\bibinfo{publisher}{Optical Society of America}, \bibinfo{year}{2017}), p.
  \bibinfo{pages}{SM3K.6}.

\bibitem[{\citenamefont{Kr\"{u}ckel et~al.}(2015)\citenamefont{Kr\"{u}ckel,
  F\"{u}l\"{o}p, Klintberg, Bengtsson, Andrekson, and
  Torres-Company}}]{Kruckel:15}
\bibinfo{author}{\bibfnamefont{C.~J.} \bibnamefont{Kr\"{u}ckel}},
  \bibinfo{author}{\bibfnamefont{A.}~\bibnamefont{F\"{u}l\"{o}p}},
  \bibinfo{author}{\bibfnamefont{T.}~\bibnamefont{Klintberg}},
  \bibinfo{author}{\bibfnamefont{J.}~\bibnamefont{Bengtsson}},
  \bibinfo{author}{\bibfnamefont{P.~A.} \bibnamefont{Andrekson}},
  \bibnamefont{and}
  \bibinfo{author}{\bibfnamefont{V.}~\bibnamefont{Torres-Company}},
  \bibinfo{journal}{Opt. Express} \textbf{\bibinfo{volume}{23}},
  \bibinfo{pages}{25827} (\bibinfo{year}{2015}).

\bibitem[{\citenamefont{Borselli et~al.}(2005)\citenamefont{Borselli, Johnson,
  and Painter}}]{Borselli:05}
\bibinfo{author}{\bibfnamefont{M.}~\bibnamefont{Borselli}},
  \bibinfo{author}{\bibfnamefont{T.~J.} \bibnamefont{Johnson}},
  \bibnamefont{and} \bibinfo{author}{\bibfnamefont{O.}~\bibnamefont{Painter}},
  \bibinfo{journal}{Opt. Express} \textbf{\bibinfo{volume}{13}},
  \bibinfo{pages}{1515} (\bibinfo{year}{2005}).

\bibitem[{\citenamefont{Porkolab et~al.}(2014)\citenamefont{Porkolab,
  Apiratikul, Wang, Guo, and Richardson}}]{Porkolab:14}
\bibinfo{author}{\bibfnamefont{G.~A.} \bibnamefont{Porkolab}},
  \bibinfo{author}{\bibfnamefont{P.}~\bibnamefont{Apiratikul}},
  \bibinfo{author}{\bibfnamefont{B.}~\bibnamefont{Wang}},
  \bibinfo{author}{\bibfnamefont{S.~H.} \bibnamefont{Guo}}, \bibnamefont{and}
  \bibinfo{author}{\bibfnamefont{C.~J.~K.} \bibnamefont{Richardson}},
  \bibinfo{journal}{Opt. Express} \textbf{\bibinfo{volume}{22}},
  \bibinfo{pages}{7733} (\bibinfo{year}{2014}).

\bibitem[{cit({\natexlab{b}})}]{cite:comsol}
\bibinfo{note}{\url{https://www.comsol.com}}.

\bibitem[{\citenamefont{Meng et~al.}(2015)\citenamefont{Meng, Lee, Dagenais,
  and Rolston}}]{meng2015nanowaveguide}
\bibinfo{author}{\bibfnamefont{Y.}~\bibnamefont{Meng}},
  \bibinfo{author}{\bibfnamefont{J.}~\bibnamefont{Lee}},
  \bibinfo{author}{\bibfnamefont{M.}~\bibnamefont{Dagenais}}, \bibnamefont{and}
  \bibinfo{author}{\bibfnamefont{S.}~\bibnamefont{Rolston}},
  \bibinfo{journal}{Applied Physics Letters} \textbf{\bibinfo{volume}{107}},
  \bibinfo{pages}{091110} (\bibinfo{year}{2015}).

\bibitem[{\citenamefont{Stievater et~al.}(2016)\citenamefont{Stievater, Kozak,
  Pruessner, Mahon, Park, Rabinovich, and Fatemi}}]{stievater2016modal}
\bibinfo{author}{\bibfnamefont{T.~H.} \bibnamefont{Stievater}},
  \bibinfo{author}{\bibfnamefont{D.~A.} \bibnamefont{Kozak}},
  \bibinfo{author}{\bibfnamefont{M.~W.} \bibnamefont{Pruessner}},
  \bibinfo{author}{\bibfnamefont{R.}~\bibnamefont{Mahon}},
  \bibinfo{author}{\bibfnamefont{D.}~\bibnamefont{Park}},
  \bibinfo{author}{\bibfnamefont{W.~S.} \bibnamefont{Rabinovich}},
  \bibnamefont{and} \bibinfo{author}{\bibfnamefont{F.~K.}
  \bibnamefont{Fatemi}}, \bibinfo{journal}{Optical Materials Express}
  \textbf{\bibinfo{volume}{6}}, \bibinfo{pages}{3826} (\bibinfo{year}{2016}).

\bibitem[{cit({\natexlab{c}})}]{cite:hlattice}
\bibinfo{note}{Alternatively, one may also excite the blue-detuned mode via
  single end of the waveguide bus using a laser of finite line width $\delta
  \nu \gg \beta$ ($\beta<2\pi \times 0.6~$GHz) so that no coherent
  back-scattering can establish within the microring}.

\bibitem[{\citenamefont{Stern et~al.}(2011)\citenamefont{Stern, Alton, and
  Kimble}}]{Stern_2011}
\bibinfo{author}{\bibfnamefont{N.~P.} \bibnamefont{Stern}},
  \bibinfo{author}{\bibfnamefont{D.~J.} \bibnamefont{Alton}}, \bibnamefont{and}
  \bibinfo{author}{\bibfnamefont{H.~J.} \bibnamefont{Kimble}},
  \bibinfo{journal}{New Journal of Physics} \textbf{\bibinfo{volume}{13}},
  \bibinfo{pages}{085004} (\bibinfo{year}{2011}),
  \urlprefix\url{https://doi.org/10.1088%2F1367-2630%2F13%2F8%2F085004}.

\bibitem[{\citenamefont{Barredo et~al.}(2016)\citenamefont{Barredo, Leseleuc,
  Lienhard, Lahaye, and Browaeys}}]{barredo_atom-by-atom_2016}
\bibinfo{author}{\bibfnamefont{D.}~\bibnamefont{Barredo}},
  \bibinfo{author}{\bibfnamefont{S.~d.} \bibnamefont{Leseleuc}},
  \bibinfo{author}{\bibfnamefont{V.}~\bibnamefont{Lienhard}},
  \bibinfo{author}{\bibfnamefont{T.}~\bibnamefont{Lahaye}}, \bibnamefont{and}
  \bibinfo{author}{\bibfnamefont{A.}~\bibnamefont{Browaeys}},
  \bibinfo{journal}{Science} \textbf{\bibinfo{volume}{354}},
  \bibinfo{pages}{1021} (\bibinfo{year}{2016}), ISSN \bibinfo{issn}{0036-8075,
  1095-9203}.

\bibitem[{\citenamefont{Hung et~al.}(2016)\citenamefont{Hung,
  Gonz{\'a}lez-Tudela, Cirac, and Kimble}}]{hung_quantum_2016}
\bibinfo{author}{\bibfnamefont{C.-L.} \bibnamefont{Hung}},
  \bibinfo{author}{\bibfnamefont{A.}~\bibnamefont{Gonz{\'a}lez-Tudela}},
  \bibinfo{author}{\bibfnamefont{J.~I.} \bibnamefont{Cirac}}, \bibnamefont{and}
  \bibinfo{author}{\bibfnamefont{H.~J.} \bibnamefont{Kimble}},
  \bibinfo{journal}{Proceedings of the National Academy of Sciences}
  \textbf{\bibinfo{volume}{113}}, \bibinfo{pages}{E4946}
  (\bibinfo{year}{2016}), ISSN \bibinfo{issn}{0027-8424, 1091-6490}.

\bibitem[{\citenamefont{Wei et~al.}(2015)\citenamefont{Wei, Stanev, Czaplewski,
  Jung, and Stern}}]{wei2015silicon}
\bibinfo{author}{\bibfnamefont{G.}~\bibnamefont{Wei}},
  \bibinfo{author}{\bibfnamefont{T.~K.} \bibnamefont{Stanev}},
  \bibinfo{author}{\bibfnamefont{D.~A.} \bibnamefont{Czaplewski}},
  \bibinfo{author}{\bibfnamefont{I.~W.} \bibnamefont{Jung}}, \bibnamefont{and}
  \bibinfo{author}{\bibfnamefont{N.~P.} \bibnamefont{Stern}},
  \bibinfo{journal}{Applied Physics Letters} \textbf{\bibinfo{volume}{107}},
  \bibinfo{pages}{091112} (\bibinfo{year}{2015}).

\bibitem[{\citenamefont{Saskin et~al.}(2019)\citenamefont{Saskin, Wilson,
  Grinkemeyer, and Thompson}}]{saskin2019narrow}
\bibinfo{author}{\bibfnamefont{S.}~\bibnamefont{Saskin}},
  \bibinfo{author}{\bibfnamefont{J.}~\bibnamefont{Wilson}},
  \bibinfo{author}{\bibfnamefont{B.}~\bibnamefont{Grinkemeyer}},
  \bibnamefont{and} \bibinfo{author}{\bibfnamefont{J.~D.}
  \bibnamefont{Thompson}}, \bibinfo{journal}{Physical review letters}
  \textbf{\bibinfo{volume}{122}}, \bibinfo{pages}{143002}
  (\bibinfo{year}{2019}).

\bibitem[{\citenamefont{Nandi et~al.}(2019)\citenamefont{Nandi, Jiang, Pak,
  Perry, Han, Bielejec, Xuan, and Hosseini}}]{nandi2019anomalous}
\bibinfo{author}{\bibfnamefont{A.}~\bibnamefont{Nandi}},
  \bibinfo{author}{\bibfnamefont{X.}~\bibnamefont{Jiang}},
  \bibinfo{author}{\bibfnamefont{D.}~\bibnamefont{Pak}},
  \bibinfo{author}{\bibfnamefont{D.}~\bibnamefont{Perry}},
  \bibinfo{author}{\bibfnamefont{K.}~\bibnamefont{Han}},
  \bibinfo{author}{\bibfnamefont{E.~S.} \bibnamefont{Bielejec}},
  \bibinfo{author}{\bibfnamefont{Y.}~\bibnamefont{Xuan}}, \bibnamefont{and}
  \bibinfo{author}{\bibfnamefont{M.}~\bibnamefont{Hosseini}},
  \bibinfo{journal}{arXiv preprint arXiv:1902.08898}  (\bibinfo{year}{2019}).

\bibitem[{\citenamefont{Ritter et~al.}(2016)\citenamefont{Ritter, Gruhler,
  Pernice, K{\"u}bler, Pfau, and L{\"o}w}}]{ritter2016coupling}
\bibinfo{author}{\bibfnamefont{R.}~\bibnamefont{Ritter}},
  \bibinfo{author}{\bibfnamefont{N.}~\bibnamefont{Gruhler}},
  \bibinfo{author}{\bibfnamefont{W.}~\bibnamefont{Pernice}},
  \bibinfo{author}{\bibfnamefont{H.}~\bibnamefont{K{\"u}bler}},
  \bibinfo{author}{\bibfnamefont{T.}~\bibnamefont{Pfau}}, \bibnamefont{and}
  \bibinfo{author}{\bibfnamefont{R.}~\bibnamefont{L{\"o}w}},
  \bibinfo{journal}{New Journal of Physics} \textbf{\bibinfo{volume}{18}},
  \bibinfo{pages}{103031} (\bibinfo{year}{2016}).

\bibitem[{\citenamefont{Ritter et~al.}(2018)\citenamefont{Ritter, Gruhler,
  Dobbertin, K{\"u}bler, Scheel, Pernice, Pfau, and
  L{\"o}w}}]{ritter2018coupling}
\bibinfo{author}{\bibfnamefont{R.}~\bibnamefont{Ritter}},
  \bibinfo{author}{\bibfnamefont{N.}~\bibnamefont{Gruhler}},
  \bibinfo{author}{\bibfnamefont{H.}~\bibnamefont{Dobbertin}},
  \bibinfo{author}{\bibfnamefont{H.}~\bibnamefont{K{\"u}bler}},
  \bibinfo{author}{\bibfnamefont{S.}~\bibnamefont{Scheel}},
  \bibinfo{author}{\bibfnamefont{W.}~\bibnamefont{Pernice}},
  \bibinfo{author}{\bibfnamefont{T.}~\bibnamefont{Pfau}}, \bibnamefont{and}
  \bibinfo{author}{\bibfnamefont{R.}~\bibnamefont{L{\"o}w}},
  \bibinfo{journal}{Physical Review X} \textbf{\bibinfo{volume}{8}},
  \bibinfo{pages}{021032} (\bibinfo{year}{2018}).

\bibitem[{\citenamefont{{Oxborrow}}(2007)}]{4230891}
\bibinfo{author}{\bibfnamefont{M.}~\bibnamefont{{Oxborrow}}},
  \bibinfo{journal}{IEEE Transactions on Microwave Theory and Techniques}
  \textbf{\bibinfo{volume}{55}}, \bibinfo{pages}{1209} (\bibinfo{year}{2007}),
  ISSN \bibinfo{issn}{0018-9480}.

\bibitem[{\citenamefont{Cheema and Kirk}(2010)}]{cheema2010implementation}
\bibinfo{author}{\bibfnamefont{M.~I.} \bibnamefont{Cheema}} \bibnamefont{and}
  \bibinfo{author}{\bibfnamefont{A.~G.} \bibnamefont{Kirk}}, in
  \emph{\bibinfo{booktitle}{COMSOL conference}} (\bibinfo{year}{2010}).

\bibitem[{\citenamefont{Pfeiffer et~al.}(2018)\citenamefont{Pfeiffer, Liu,
  Raja, Morais, Ghadiani, and Kippenberg}}]{Pfeiffer:18}
\bibinfo{author}{\bibfnamefont{M.~H.~P.} \bibnamefont{Pfeiffer}},
  \bibinfo{author}{\bibfnamefont{J.}~\bibnamefont{Liu}},
  \bibinfo{author}{\bibfnamefont{A.~S.} \bibnamefont{Raja}},
  \bibinfo{author}{\bibfnamefont{T.}~\bibnamefont{Morais}},
  \bibinfo{author}{\bibfnamefont{B.}~\bibnamefont{Ghadiani}}, \bibnamefont{and}
  \bibinfo{author}{\bibfnamefont{T.~J.} \bibnamefont{Kippenberg}},
  \bibinfo{journal}{Optica} \textbf{\bibinfo{volume}{5}}, \bibinfo{pages}{884}
  (\bibinfo{year}{2018}).

\bibitem[{\citenamefont{Ding et~al.}(2012)\citenamefont{Ding, Goban, Choi, and
  Kimble}}]{ding2012corrections}
\bibinfo{author}{\bibfnamefont{D.}~\bibnamefont{Ding}},
  \bibinfo{author}{\bibfnamefont{A.}~\bibnamefont{Goban}},
  \bibinfo{author}{\bibfnamefont{K.}~\bibnamefont{Choi}}, \bibnamefont{and}
  \bibinfo{author}{\bibfnamefont{H.}~\bibnamefont{Kimble}},
  \bibinfo{journal}{arXiv preprint arXiv:1212.4941}  (\bibinfo{year}{2012}).

\bibitem[{\citenamefont{Payne and Lacey}(1994)}]{payne1994theoretical}
\bibinfo{author}{\bibfnamefont{F.}~\bibnamefont{Payne}} \bibnamefont{and}
  \bibinfo{author}{\bibfnamefont{J.}~\bibnamefont{Lacey}},
  \bibinfo{journal}{Optical and Quantum Electronics}
  \textbf{\bibinfo{volume}{26}}, \bibinfo{pages}{977} (\bibinfo{year}{1994}).

\bibitem[{\citenamefont{Poulton et~al.}(2006)\citenamefont{Poulton, Koos,
  Fujii, Pfrang, Schimmel, Leuthold, and Freude}}]{poulton2006radiation}
\bibinfo{author}{\bibfnamefont{C.~G.} \bibnamefont{Poulton}},
  \bibinfo{author}{\bibfnamefont{C.}~\bibnamefont{Koos}},
  \bibinfo{author}{\bibfnamefont{M.}~\bibnamefont{Fujii}},
  \bibinfo{author}{\bibfnamefont{A.}~\bibnamefont{Pfrang}},
  \bibinfo{author}{\bibfnamefont{T.}~\bibnamefont{Schimmel}},
  \bibinfo{author}{\bibfnamefont{J.}~\bibnamefont{Leuthold}}, \bibnamefont{and}
  \bibinfo{author}{\bibfnamefont{W.}~\bibnamefont{Freude}},
  \bibinfo{journal}{IEEE Journal of selected topics in quantum electronics}
  \textbf{\bibinfo{volume}{12}}, \bibinfo{pages}{1306} (\bibinfo{year}{2006}).

\bibitem[{\citenamefont{{Kuznetsov}}(1985)}]{1074216}
\bibinfo{author}{\bibfnamefont{M.}~\bibnamefont{{Kuznetsov}}},
  \bibinfo{journal}{Journal of Lightwave Technology}
  \textbf{\bibinfo{volume}{3}}, \bibinfo{pages}{674} (\bibinfo{year}{1985}),
  ISSN \bibinfo{issn}{0733-8724}.

\end{thebibliography}
